\documentclass[opre,nonblindrev]{informs3} 

\OneAndAHalfSpacedXI 

\usepackage{endnotes}
\let\enotesize=\normalsize
\def\notesname{Endnotes}%
\def\makeenmark{$^{\theenmark}$}
\def\enoteformat{\rightskip0pt\leftskip0pt\parindent=1.75em
  \leavevmode\llap{\theenmark.\enskip}}

\usepackage{algorithm, algorithmic}
\usepackage{latexsym, amsmath, amssymb, color, colortbl, rotating, multirow, hyperref, enumitem}
\usepackage[normalem]{ulem}

\usepackage{booktabs,multirow}
\usepackage{bm}
\usepackage{bbm}
\usepackage{relsize}

\usepackage{verbatim, csquotes, mathrsfs}
\usepackage{xcolor}
\newcommand{\set}[1]{\left\{#1\right\}}

\newcommand{\B}{\mathcal{B}}
\newcommand{\F}{\mathcal{F}}

\newcommand{\calD}{\mathcal{D}}

\newcommand{\Pcal}{\mathcal{P}}
\newcommand{\pci}{\B^{(i)}}

\newcommand{\norm}[1]{\left\Vert#1\right\Vert}
\newcommand{\abs}[1]{\left\vert#1\right\vert}
\newcommand{\expt}[1]{\bigg[#1\bigg]}
\newcommand{\bexpt}[1]{\Bigg[#1\Bigg]}
\newcommand{\bmx}{\bm{x}}

\newcommand{\bmy}{\bm{y}}
\newcommand{\bmz}{\bm{Z}}
\newcommand{\bmX}{\bm{X}}
\newcommand{\bmU}{\bm{U}}
\newcommand{\bmV}{\bm{V}}
\newcommand{\bmY}{\bm{Y}}
\newcommand{\bmW}{\bm{W}}
\newcommand{\bmH}{\bm{H}}
\newcommand{\ind}{\mathbf{1}}
\newcommand{\E}{\mathbb{E}}
\newcommand{\I}{\mathcal{I}}

\newcommand{\Real}{\mathbb{R}}
\newcommand{\bin}{\set{-1,+1}}

\newcommand{\mi}[1]{{#1}^{(i)}}
\newcommand{\alphak}{\alpha_k}

\newcommand{\xij}{\bm{X}_{ij}}

\newcommand{\hbmz}{\hat{\bm{I}}}

\newcommand{\djplus}{d_j^{\text{purch}}}
\newcommand{\djminus}{d_j^{\text{non-click}}}
\newcommand{\djzero}{d_j^{\text{click}}}
\newcommand{\xiplus}{\bm{X}_i^+}
\newcommand{\ximinus}{\bm{X}_i^-}
\newcommand{\xibplus}{\bm{X}_{ib}^+}
\newcommand{\xibpplus}{\bm{X}_{i'b}^+}
\newcommand{\fbplus}{\bm{F}_{0b}^+}
\newcommand{\fibplus}{\bm{F}_{ib}^+}

\newcommand{\xipplus}{\bm{X}_{i'}^+}

\newcommand{\cond}{\, \Big\vert \,}

\newcommand{\alphap}{\bm{F}_0^+}

\newcommand{\pool}{\mathrm{pool}}
\newcommand{\fpool}{f_{\pool}}
\newcommand{\alphab}{\alpha}
\newcommand{\alphavec}{\bm{\alpha}}
\newcommand{\alphapool}{\alpha_{\rm pool}}
\newcommand{\alphabpool}{\bar{\alpha}_{\rm pool}}
\newcommand{\alphacatpool}[1]{\alpha_{#1, {\rm pool}}}
\newcommand{\amin}{\alpha_{\min}}
\newcommand{\hatalphapool}{\hat{\alpha}_{\rm pool}}
\newcommand{\baralphabpool}{\bar{\alpha}_{b, {\rm pool}}}

\newcommand{\catcons}{\Gamma}

\makeatletter
\def\moverlay{\mathpalette\mov@rlay}
\def\mov@rlay#1#2{\leavevmode\vtop{%
   \baselineskip\z@skip \lineskiplimit-\maxdimen
   \ialign{\hfil$\m@th#1##$\hfil\cr#2\crcr}}}
\newcommand{\charfusion}[3][\mathord]{
    #1{\ifx#1\mathop\vphantom{#2}\fi
        \mathpalette\mov@rlay{#2\cr#3}
      }
    \ifx#1\mathop\expandafter\displaylimits\fi}
\makeatother
\makeatletter
\def\mathcolor#1#{\@mathcolor{#1}}
\def\@mathcolor#1#2#3{%
  \protect\leavevmode
  \begingroup
    \color#1{#2}#3%
  \endgroup
}
\makeatother

\newcommand{\cupdot}{\charfusion[\mathbin]{\cup}{\cdot}}

\newcommand{\defas}{\overset{\mathrm{def}}{=}}

\newcommand{\proj}{\mathrm{pscore}}
\newcommand{\projrv}{\bm{\mathrm{PSCORE}}}
\newcommand{\projrvec}{\overrightarrow{\bm{\mathrm{PSCORE}}}}

\newcommand{\xobs}{\bm{x}^{\obs}}
\newcommand{\xmis}{\bm{x}^{\mis}}
\newcommand{\projmat}{\bm{\mathrm{PSCORES}}}

\usepackage{natbib}
 \bibpunct[, ]{(}{)}{,}{a}{}{,}%
 \def\bibfont{\small}%
 \def\bibsep{\smallskipamount}%
 \def\bibhang{24pt}%
\TheoremsNumberedThrough     
\ECRepeatTheorems

\EquationsNumberedThrough    

\begin{document}


\RUNAUTHOR{Jagabathula, Subramanian and Venkataraman}

\RUNTITLE{A Model-based Projection Technique for Segmenting Customers}

\TITLE{{\sf A Model-based Projection Technique for Segmenting Customers}}

\ARTICLEAUTHORS{%
\AUTHOR{Srikanth Jagabathula}
\AFF{Stern School of Business, New York University, New York, NY 10012,\EMAIL{sjagabat@stern.nyu.edu}} 
\AUTHOR{Lakshminarayanan Subramanian, Ashwin Venkataraman}
\AFF{Courant Institute of Mathematical Sciences, New York University, New York, NY 10012, \EMAIL{\{lakshmi,ashwin\}@cs.nyu.edu}}
} 

\ABSTRACT{%
We consider the problem of segmenting a large population of customers into non-overlapping groups with similar preferences, using diverse preference observations such as purchases, ratings, clicks, etc.\ over subsets of items. We focus on the setting where the universe of items is large (ranging from thousands to millions) and unstructured (lacking well-defined attributes) and each customer provides observations for only a few items. These data characteristics limit the applicability of existing techniques in marketing and machine learning. To overcome these limitations, we propose a {\em model-based projection technique}, which transforms the diverse set of observations into a more comparable scale and deals with missing data by projecting the transformed data onto a low-dimensional space. We then cluster the projected data to obtain the customer segments. Theoretically, we derive precise necessary and sufficient conditions that guarantee asymptotic recovery of the true customer segments. Empirically, we demonstrate the speed and performance of our method in two real-world case studies: (a) 84\% improvement in the accuracy of new movie recommendations on the MovieLens data set and (b) 6\% improvement in the performance of similar item recommendations algorithm on an offline dataset at eBay. We show that our method outperforms standard latent-class and demographic-based techniques.
}%


\maketitle

%

\vspace{-1.5em}
\section{Introduction}
\label{sec:intro}

Customer segmentation is a key marketing problem faced by a firm. It involves grouping customers into non-overlapping segments
with each segment comprising customers having similar needs and preferences. Accurately segmenting its customers allows a firm to effectively customize its product offerings, promotions, and recommendations to the particular preferences of its customers~\citep{smith1956product}. This is particularly the case when firms do not have sufficient number of observations per customer to accurately personalize their decisions to individual customers. For example, Netflix and eBay.com help their customers navigate their large catalogs by offering new movie and similar item recommendations, respectively. However, new movies on Netflix lack ratings or viewing data\footnote{This problem is popularly referred to as the `cold-start' problem within the recommendation systems literature.} whereas each customer might interact with only a small fraction of eBay's catalog; for example, in our sample data set, a customer on average interacted with 5 out of 2M items (see Section~\ref{sec:ebay}). As a result, individual-level personalization is not practical and customizing the recommendations to each segment can greatly increase the accuracy of recommendations.


Because firms are able to collect large amounts of data about their customers, we focus on the setting where the firm has collected fine-grained observations such as direct purchases, ratings, and clicks over a subset of items, in addition to any demographic data such as age, gender, income, etc., for each customer. Our goal is to use these data to classify a large population of customers into corresponding segments in order to improve the accuracy of a given prediction task. The observations are collected over a universe of items that is unstructured, lacking well-defined feature information, and large, consisting of thousands to millions of items. The data may comprise information on diverse types of actions such as purchases, clicks, ratings, etc., which are represented on different scales. Moreover, the observations from each customer are highly incomplete and span only a small fraction of the entire item universe. Such data characteristics are common in practice. For instance, the eBay marketplace offers a diverse product catalog consisting of products ranging from a Fitbit tracker/iPhone (products with well-defined attributes) to obscure antiques and collectibles, which are highly unstructured. Of these, each customer may purchase/click/rate only a few items.

The literature in marketing and machine learning has studied the problem of segmentation but the above data characteristics preclude the applicability of existing techniques. Specifically, most techniques within marketing focus on characterizing the market segments in terms of product and customer features by analyzing structured products and small samples of customer populations; consequently, they do not scale to directly classifying a large population of customers. The standard clustering techniques in machine learning (see~\citealp{jain2010data} for a review), on the other hand, are designed for such direct classification and rely on a similarity measure used to determine if two customers should belong to the same segment or not. However, the diversity and incompleteness of observations make it challenging to construct a meaningful similarity measure. For example, it is difficult to assess the degree of similarity between a customer who purchased an item and another who has rated the same item, or between two customers who have purchased completely non-overlapping sets of products.

To overcome the above limitations, we propose a {\em model-based projection technique} that extends the extant clustering techniques in machine learning to handle categorical observations from diverse data sources and having (many) missing entries. The algorithm takes as inputs the observations from a large population of customers and a probabilistic model class describing how the observations are generated from an individual customer. The choice of the model class is determined by the prediction task at hand, as described below, and provides a systematic way to incorporate domain knowledge by leveraging the existing literature in marketing, which has proposed rich models describing individual customer behavior. It outputs a representation of each customer as a vector in a low-dimensional Euclidean space whose dimension is much smaller than the number of items in the universe. The vector representations are then clustered, using a standard technique such as the $k$-means algorithm, to obtain the corresponding segments. In particular, the algorithm proceeds in two sequential steps: {\em transform} and {\em project}. The {\em transform} step is designed to address the issue of diversity of the observed signals by transforming the categorical observations into a continuous scale that makes different observations (such as purchases and ratings) comparable. The {\em project} step then deals with the issue of missing data by projecting the transformed observations onto a low-dimensional space, to obtain a vector representation for each customer.

The key novelty of our algorithm is the {\em transform} step. 
This step uses a probabilistic model to convert a categorical observation into the corresponding (log-)likelihood of the observation under the model. For example, if the probability that an item is liked by a customer is given by $\alpha \in [0, 1]$, then a ``like'' observation is transformed into $\log \alpha$ and a ``dislike'' observation is transformed into $\log(1- \alpha)$. We call our algorithm {\em model-based} because the transform step relies on a probabilistic model;
Section~\ref{sec:movielens} presents a case study in which we illustrate the choice of the model when the objective is to accurately predict if a new movie will be liked by a customer. We estimate the model parameters by pooling together the data from all customers and ignoring the possibility that different customers may have different model parameters. This results in a model that describes a `pooled' customer---a virtual customer whose preferences reflect the aggregated preferences of the population. The likelihood transformations then measure how much a particular customer's preferences differ from those of the population's. The discussion in Section~\ref{sec:results} (see Lemmas 1, 2 and Theorems 1, 2, 3 and 4) shows that under reasonable assumptions, customers from different segments will have different (log-)likelihood values under the pooled model---allowing us to separate them out.

Our algorithm is inspired by existing ideas for clustering in the theoretical computer science literature and systematically generalizes algorithms that are popular within the machine learning community. In particular, when the customer observations are continuous and there are no missing entries, then the transform step reduces to the trivial identity mapping and our algorithm reduces to the standard spectral projection technique~\citep{vempala2004spectral,achlioptas2005spectral,kannan2005spectral} for clustering the observations from a mixture of multivariate Gaussians. When the observations are continuous but there are missing entries, our algorithm becomes a generalized matrix factorization technique---commonly used in collaborative filtering applications~\citep{rennie2005fast,mnih2007probabilistic,koren2009matrix}. 

Our work makes the following key contributions:
\begin{enumerate}
    \item {\em Novel segmentation algorithm.} Our algorithm is designed to operate on large customer populations and large collections of unstructured items. Moreover, it is (a) {\em principled:} reduces to standard algorithms in machine learning in special cases; (b) {\em fast:} order of magnitude faster than benchmark latent-class models because it requires fitting only one model (as opposed to a mixture model);
and (c) {\em flexible:} allows practitioners to systematically incorporate problem-dependent structures via model specification---providing a way to take advantage of the literature in marketing that has proposed rich models describing individual customer behavior.

    \item {\em Analytical results.} Under standard assumptions on the customer preference heterogeneity, we derive necessary and sufficient conditions for exact recovery of the true segments. Specifically, we bound the asymptotic {\em misclassification rate}, i.e. the expected fraction of customers incorrectly classified, of a nearest-neighbor classifier
which uses the customer representations obtained after the {\em project} step in our algorithm above. Given a total of $n$ observations such that each customer provides atleast $\log n$ observations, we show that the misclassification rate scales as $O\left(n^{-\frac{2\Lambda^2\amin^2}{81}}\right)$ where $0 < \amin, \Lambda < 1$ are constants that depend on the underlying parameters of the latent class model. In other words, our algorithm correctly classifies {\em all} customers into their respective segments as $n \to \infty$. 
Our results are similar in spirit to the conditions derived in the existing literature for Gaussian mixture models~\citep{kannan2005spectral}. However, existing proof techniques don't generalize to our setting. Our results are one of the first to provide such guarantees for latent-class preference models.

    \item {\em Empirical results.} We conducted three numerical studies to validate our methodology:
    \begin{itemize}
        \item {\em Faster and more accurate than EM.} Using synthetic data, we show that our method obtains more accurate segments, while being upto $11\times$ faster, than the standard EM-based latent class benchmark.
        \item {\em Cold-start problem in the {\tt MovieLens} data set.} We apply our method to the problem of recommending new movies to users, popularly referred to as the cold-start problem. On the publicly available {\tt MovieLens} dataset, we show that segmenting users via our method and customizing recommendations to each segment improves the recommendation accuracy by 48\%, 55\%, and 84\% for drama, comedy, and action genres, respectively, over a baseline population-level method that treats all users as having the same~preferences. In addition, it also outperforms the standard EM-based latent class benchmark by $8\%$, $13\%$ and $10\%$ respectively while achieving a $20\times$ speedup in the computation time.
        \item {\em Similar item recommendations on eBay.com.} We describe a real-world implementation of our segmentation methodology for personalizing similar item recommendations on eBay. The study shows that (1) segmenting the user population based on just their viewing/click/purchase activity using our approach results in upto 6\% improvement in the recommendation quality when compared to treating the population as homogeneous and (2) our algorithm can scale to large datasets. The improvement of 6\% is non-trivial because before our method, eBay tried several natural segmentations (by similarity of demographics, frequency/recency of purchases, etc.), but all of them resulted in $< 1\%$ improvement.
    \end{itemize}
      

\end{enumerate}
\subsection{Relevant literature} 
\label{sec:related_work}

Our work has connections to literature in both marketing and machine learning. We start with positioning our work in the context of segmentation techniques in the marketing literature. Customer segmentation is a classical marketing problem, with origins dating back to the work of~\cite{smith1956product}. Marketers classify various segmentation techniques into {\em a priori} versus
{\em post hoc} and {\em descriptive} versus {\em predictive} methods, giving rise to a 2 x 2 classification matrix of these techniques~\citep{wedel2012market}. Our algorithm is closest to the post-hoc predictive methods, which identify customer segments on the basis of the estimated relationship between a dependent variable and a set of predictors. 
These methods consist of clustering techniques and latent class models. The traditional method for predictive clustering is automatic interaction detection (AID), which splits the customer population into non-overlapping groups that differ maximally according to a dependent variable, such as purchase behavior, on the basis of a set of independent variables, like socioeconomic and demographic characteristics~\citep{assael1970segmenting,kass1980exploratory,maclachlan1981market}. However, these approaches typically require large sample sizes to achieve satisfactory results.
~\citet{ogawa1987approach} and~\citet{kamakura1988least} proposed hierarchical segmentation techniques tailored to conjoint analysis, which group customers such that the accuracy with which preferences/choices are predicted from product attributes or profiles is maximized.
These methods estimate parameters at the individual-level, and therefore are restricted by the number of observations available for each customer. Clusterwise regression methods overcome this limitation, as they cluster customers such that the regression fit is optimized within each cluster. The applicability of these methods to market segmentation was identified by~\cite{desarbo1989simulated} and~\cite{wedel1989consumer}, and extended by~\cite{wedel1989fuzzy} to handle partial membership of customers in segments. 

Latent class (or mixture) methods offer a statistical approach to the segmentation problem, and belong to two types: mixture regression and mixture multidimensional scaling models. Mixture regression models~\citep{wedel1994review} simultaneously group subjects into unobserved
segments and estimate a regression model within each segment, and were pioneered by~\citet{kamakura1989probabilistic} who propose a clusterwise logit model to segment households based on brand preferences and price sensitivities. This was extended by~\citet{gupta1994using} who incorporated demographic variables and~\citet{kamakura1996modeling} who incorporated differences in customer choice-making processes, resulting in models that produces identifiable and actionable segments.  
Mixture multidimensional scaling (MDS) models simultaneously estimate market segments as well as preference structures of customers in each segment, for instance, a brand map depicting the positions of the different brands on a set of unobserved dimensions assumed to influence perceptual or preference judgments of customers. See~\cite{desarbo1994latent} for a review of these methods.

One issue with the latent class approach is the discrete model of heterogeneity and it has been argued~\citep{allenby1998heterogeneity,wedel1999discrete} that individual-level response parameters are required for direct marketing approaches. As a result, continuous mixing distributions have been proposed that capture fine-grained customer heterogeneity, leading to hierarchical Bayesian estimation of models~\citep{allenby1995using,rossi1996value,allenby1998marketing}. Computation of the posterior estimates in Bayesian models, however, is typically intractable and Markov Chain Monte Carlo based techniques are employed, which entail a number of computational issues~\citep{gelman1992inference}.

The purpose of the above model-based approaches to segmenting customers is fundamentally different from our approach. These methods
focus on characterizing the market segments in terms of product and customer features (such as prices, brands, demographics, etc.) by analyzing structured products (i.e. having well-defined attributes) and small samples of customer populations; consequently, they do not scale to directly classifying a large population of customers.
Our algorithm is explicitly designed to classify the entire population of customers into segments as accurately as possible, and can be applied even when the data is less-structured or unstructured (refer to the case study in Section~\ref{sec:ebay}). Another distinction is that we can provide necessary and sufficient conditions under which our algorithm  {\em guarantees} asymptotic recovery of the true segments in a latent class model, which is unlike most prior work in the literature. In addition, our algorithm can still incorporate domain knowledge by leveraging the rich models describing customer behavior specified in existing marketing literature.

Our work also has methodological connections to literature in theoretical computer science and machine learning. Specifically, our model-based projection technique extends existing techniques for clustering real-valued observations with no missing entries~\citep{vempala2004spectral,achlioptas2005spectral,kannan2005spectral} to handle diverse categorical observations having (many) missing entries.
The {\em project} step in our segmentation algorithm has connections to matrix factorization techniques in collaborative filtering, and we point out the relevant details in our discussion of the algorithm in Section~\ref{sec:setup}.

\section{Setup and Segmentation Algorithm}
\label{sec:setup}
\newcommand{\obs}{\mathrm{obs}}
\newcommand{\mis}{\mathrm{mis}}

\newcommand{\xiobs}{\bm{x}_i^{\obs}}
\newcommand{\ximis}{\bm{x}_i^{\mis}}
\newcommand{\missing}{\phi}

Our goal is to segment a population $[m] \defas \set{1, 2, \dotsc, m}$ of $m$ customers comprised of a fixed but unknown number $K$ of non-overlapping segments. To carry out the segmentation, we assume access to individual-level observations that capture differences among the segments. The observations may come from diverse sources---organically generated clicks or purchases during online customer visits; ratings provided on review websites (such as Yelp, TripAdvisor, etc.) or recommendation systems (such as Amazon, Netflix, etc.); purchase attitudes and preferences collected from a conjoint study; and demographics such as age, gender, income, education, etc. Such data are routinely collected by firms as customers interact through various touch points. Without loss of generality, we assume that all the observations are categorical---any continuous observations may be appropriately quantized. The data sources may be coarsely curated based on the specific application but we don't assume access to fine-grained feature information.

To deal with observations from diverse sources, we consider a unified representation where each observation is mapped to a categorical label for a particular ``item'' belonging to the universe $[n] \defas \set{1, 2, \dotsc, n}$ of all items. We use the term ``item'' generically to mean different entities in different contexts. For example, when observations are product purchases, the items are products and the labels binary purchase/no-purchase signals. When observations are choices from a collection of offer sets (such as those collected in a choice-based conjoint study), the items are offer sets and labels the IDs of chosen products. Finally, when observations are ratings for movies, the items are movies and the labels star ratings. Therefore, our representation provides a compact and general way to capture diverse signals. We index a typical customer by $i$, item by $j$, and segment by $k$.


In practice, we observe labels for only a small subset of the items for each customer. Because the numbers of observations can widely differ across customers, we represent the observed labels using an edge-labeled bipartite graph $\Pcal$, defined between the customers and the items. An edge $(i,j)$ denotes that we have observed a label from customer $i$ for item $j$, with the edge-label $x_{ij}$ representing the observed label. We call this graph the {\em customer-item preference graph}. We let $\bm{x}_i$ denote the vector\footnote{We use bold lower-case letters like $\bm{x}, \bm{y}$ etc. to represent vectors} of observations for customer $i$ with $x_{ij} = \missing$ if the label for item $j$ from customer $i$ is unobserved/missing. Let $N(i)$ denote the set of items for which we have observations for customer $i$. It follows from our definitions that $N(i)$ also denotes the set of neighbors of the customer node $i$ in the bipartite graph $\Pcal$ and the degree $d_i \defas \abs{N(i)}$, the size of the set $N(i)$, denotes the number of observations for customer $i$. Note that the observations for each customer are typically highly incomplete and therefore, $d_i \ll n$ and the bipartite graph $\Pcal$ is highly sparse. 

 
In order to carry out the segmentation, we assume that different segments are characterized by different preference parameters and a {\em model} relates the latent parameters to the observations. Specifically, the customer labels are generated according to a parametric model from the model-class $\F(\Omega) = \set{f(\bmx; \omega) \; \colon \; \omega \in \Omega}$, where $\Omega$ is the parameter space that indexes the models $f \in \F$; $\bmx = (x_1,x_2,\ldots,x_n) \in \B := \mathcal{X}_1 \times \dotsb \times \mathcal{X}_n$ is the vector of item labels, and $f(\bmx; \omega)$ is the probability of observing the item labels $\bmx$ from a customer with model parameter $\omega$. Here, $\mathcal{X}_j$ is the domain of possible categorical labels for item $j$. When all the labels are binary, $\mathcal{X}_j = \set{0, 1}$ and $\B = \set{0, 1}^n$. The choice of the parametric family depends on the application context and the prediction problem at hand. For instance, if the task is to predict whether a particular segment of customers likes a movie or not, then $\F$ can be chosen to be the binary logit model class. Instead, if the task is to predict the movie rating (say, on a 5-star scale) for each segment, then $\F$ can be the ordered logit model class. Finally, if the task is to predict which item each segment will purchase, then $\F$ can be the multinomial logit (MNL) model class. Depending on the application, other models proposed within the marketing literature may be used.  We provide a concrete illustration as part of a case study with the MovieLens dataset, described in Section~\ref{sec:movielens}. 

A population comprised of $K$ segments is described by $K$ distinct models $f_1,f_2,\dotsb,f_K$ with corresponding parameters $\omega_1,\omega_2,\dotsb, \omega_K$ so that the observations from customers in segment $k$ are generated according to model $f_k$. Specifically, we assume that customer $i$ in segment $k$ generates the label vector $\bmx_i \sim f_k$ and we observe the labels $x_{ij}$ for all the items $j \in N(i)$, for some preference graph $\Pcal$. Further, for ease of notation, we drop the explicit dependence of models in $\F$ on the parameter $\omega$ in the remainder of the discussion. Let $\bm{x}_i^{\obs} \defas \left( x_{ij} \right)_{j \in N(i)}$ denote the observed label vector from customer $i$ and define the domain $\pci = \set{(x_j)_{j \in N(i)} \; | \; \bmx \in \B}$. Given any model $f \in \F$, we define $\mi{f}(\bmy) \defas \sum_{\bm{x}_i^{\mis}} f(\bmy, \bm{x}_i^{\mis})$ for each $\bmy \in \pci$, where $\xmis_i$ represent the missing labels vector for customer $i$ and the summation is over all possible feasible missing label vectors when given the observations $\bm{y}$. Observe that $\mi{f}$ defines a distribution over $\pci$. Finally, let $\abs{\xiobs}$ denote the length of the vector $\xiobs$; we have $\abs{\xiobs} = \abs{N(i)}$.

We now describe our segmentation algorithm. For purposes of exposition, we describe three increasingly sophisticated variants of our algorithm. The first variant assumes that the number of observations is the same across all customers---that is, all customers have the same degree $\ell$ in the preference graph $\Pcal$---and the model class $\F$ is completely specified. The second variant relaxes the equal degree assumption, and the third variant allows for partial specification of the model class~$\F$.

\subsection{Segmentation algorithm for $\ell$-regular preference graph $\Pcal$}
\label{sec:lreg_graph}
We first focus on the case when we have the same number $\ell$ of observations for all the customers, so that $d_i = \ell$ for all $i$, resulting in an $\ell$-regular preference graph $\Pcal$. We describe the algorithm assuming the number of segments $K$ is specified and discuss below how $K$ may be estimated from the data. The precise description is presented in Algorithm~\ref{alg:lreg_seg}. The algorithm takes as inputs observations in the form of the preference graph $\mathcal{P}$ and the model family $\mathcal{F}$ and outputs a uni-dimensional representation of each customer. The algorithm proceeds as follows. It starts with the hypothesis that the population of customers is in fact homogeneous and looks for evidence of heterogeneity to refute the hypothesis. Under this hypothesis, it follows that the $m$ observations $\xobs_1, \xobs_2, \dotsc, \xobs_m$ are i.i.d.\ samples generated according to a single model in $\F$. Therefore, the algorithm estimates the parameters of a `pooled' model $f_{\pool} \in \F$ by pooling together all the observations and using a standard technique such as the maximum likelihood estimation (MLE) method. 
As a concrete example, consider the task of predicting whether a segment of customers likes a movie or not, so that $\F$ is chosen to be the binary logit model class where movie $j$ is liked with probability $e^{\omega_j}/(1 + e^{\omega_j})$, independent of the other movies. Then, the parameters of the pooled model can be estimated by solving the following MLE problem:
\[ \max_{\omega_1,\omega_2,\ldots,\omega_n} \sum\limits_{i=1}^m \sum\limits_{j \in N(i)} \log \left(\frac{e^{\ind[x_{ij} = +1] \cdot \omega_j}}{1 + e^{\omega_j}}\right), \text{ where } x_{ij} = +1 \text{ if customer $i$ likes movie $j$ and $-1$ for dislike} \]
Because the objective function is separable, the optimal solution can be shown to be given by $\hat{\omega}_j = \log\left( \frac{\sum_{i : j \in N(i)} \ind[x_{ij} = +1]}{\sum_{i : j \in N(i)} \ind[x_{ij} = -1]} \right)$.

Once the pooled model is estimated, the algorithm assesses if the hypothesis holds by checking how well the pooled model explains the observed customer labels. Specifically, it quantifies the model fit by computing the (normalized) negative log-likelihood of observing $\xiobs$ under the pooled model, i.e., $\proj_i \defas \frac{1}{d_i} \cdot \left(-\log \mi{f}_{\pool}(\xiobs)\right)$. A large value of $\proj_i$ indicates that the observation $\xiobs$ is not well explained by the pooled model or that customer $i$'s preferences are ``far away'' from that of the population. We term $\proj_i$ the {\em model-based projection score}, or simply the projection score, for customer $i$ because it is obtained by ``projecting'' categorical observations ($\xiobs$) into real numbers by means of a model ($f_{\pool}$). Note that the projection scores yield one-dimensional representations of the customers.
The entire process is summarized in Algorithm~\ref{alg:lreg_seg}.

\begin{algorithm}
\caption{Segmentation algorithm with degree normalization}
\label{alg:lreg_seg}
\begin{algorithmic}[1]
\STATE \textbf{Input:} observed labels $\xobs_1, \xobs_2, \ldots, \xobs_m$ where $\abs{\xiobs} = d_i~\forall~i$, model class $\F$
\STATE $\fpool \gets$ estimated pooled model in family $\F$
\STATE For each customer $i$ with observation $\xiobs$,
compute the projection score: \\
$\proj_i = \frac{1}{d_i} \cdot \left(-\log \mi{f}_{\pool}(\xiobs) \right)$
\STATE \textbf{Output:} $\set{\proj_1,\proj_2,\ldots,\proj_m}$
\end{algorithmic}
\end{algorithm}
The projection scores obtained by the algorithm are then clustered into $K$ groups using a standard distance-based clustering technique, such as the $k$-means algorithm, to recover the customer segments.
We discuss how to estimate $K$ at the end of this section.

We make the following remarks. First, our model-based projection technique is inspired by the classical statistical technique of analysis-of-variance (ANOVA), which tests the hypothesis of whether a collection of samples are generated from the same underlying population or not. For that, the test starts with the null hypothesis that there is no heterogeneity, fits a single model by pooling together all the data, and then computes the likelihood of the observations under the pooled model. If the likelihood is low (i.e., below a threshold), the test rejects the null hypothesis and concludes that the samples come from different populations. Our algorithm essentially separates customers based on the heterogeneity within these likelihood values.

Second, to understand why our algorithm should be able to separate the segments, consider the following simple case. Suppose a customer from segment $k$ likes any item $j$ with probability $f_k(\text{like}) = \alpha_k$ and dislikes it with probability $f_k(\text{dislike}) = 1 - \alpha_k$ for some $\alpha_k \in [0, 1]$. Different segments differ on the value of the parameter $\alpha_k$. Suppose $q_k$ denotes the size of segment $k$, where $\sum_k q_k = 1$ and $q_k > 0$ for all $k$. Now, when we pool together a large number of observations from these customers, we should essentially observe that the population as a whole likes an item with probability $f_{\pool}(\text{like}) \defas \sum_k q_k \alpha_k$; this corresponds to the pooled model. Under the pooled model, we obtain the projection score for customer $i$ as $\frac{1}{\abs{N(i)}} \sum_{j \in N(i)} - \log f_{\pool}(x_{ij})$ where each $x_{ij} \in \set{\text{like}, \text{dislike}}$. Now assuming that $\abs{N(i)}$ is large and because the $x_{ij}$'s are randomly generated, the projection score should essentially concentrate around the expectation $\E_{\bm{X}_i \sim f_k}[-\log f_{\pool}(\bm{X}_{i})]$ where r.v.\ $\bm{X}_{i}$ takes value ``like'' with probability $\alpha_k$ and ``dislike'' with probability $1 - \alpha_k$, when customer $i$ belongs to segment $k$. The value $\E_{\bm{X}_i \sim f_k}[-\log f_{\pool}(\bm{X}_{i})]$ is the cross-entropy between the distributions $f_k$ and $f_{\pool}$. Therefore, if the cross-entropies for the different segments are distinct, our algorithm should be able to separate the segments.\footnote{Note that the cross-entropy is not a distance measure between distributions unlike the standard KL (Kullback-Leibler) divergence. Consequently, even when $f_k = f_{\pool}$ for some segment $k$, the cross-entropy is not zero. Our algorithm relies on the cross-entropies being distinct to recover the underlying segments} We formalize and generalize these arguments in~Section~\ref{sec:results}.

Third, our algorithm fits only one model---the `pooled' model---unlike a classical latent class approach that fits, typically using the Expectation-Maximization (EM) method, a mixture distribution $g(\bmx) = \sum_{k} q_k f_k(\bmx)$, where all customers in segment $k$ are described by model $f_k \in \F$ and $q_k$ represents the size (or proportion) of segment $k$. This affords our algorithm two advantages: (a) {\em speed:} up to $11 \times$ faster than the standard latent class benchmark (see~Section~\ref{sec:simul_exp}) without the issues of initialization and convergence, typical of EM-methods; and (b) {\em flexibility:} allows for fitting models from complex parametric families $\F$, that more closely explain the customer observations. 

\vspace{10pt}
\noindent{\em Estimating the number of segments}. Although our algorithm description assumed that the number of customer segments $K$ is provided as input, it can actually provide data-driven guidance on how to pick $K$---which is often unknown in practice. While existing methods rely on cross-validation and information-theoretic measures such as AIC, BIC etc. (see~\citealp{mclachlan2004finite} for details), our algorithm can also rely on the structure of the projection scores to estimate $K$. As argued above, when there are $K$ underlying segments in the population, the projection scores will concentrate around $K$ distinct values corresponding to the cross-entropies between the distributions $f_k$ and $f_{\pool}$. Therefore, when plotted on a line, these scores will separate into clusters, with the number of clusters corresponding to the number of segments. In practice, however, the scores may not cleanly separate out. Therefore, we use the following systematic procedure: estimate the density of the customer projection scores $\set{\proj_i}_{i \in [m]}$ using a general purpose technique such as Kernel Density Estimation (KDE) and associate a segment with each peak (i.e. local maximum) in the density. The estimated density should capture the underlying clustering structure with each peak in the density corresponding to the value around which many of the projection scores concentrate and as a result, should be able to recover the underlying number of segments. We use this technique to estimate the number of segments in our real-world case studies (see Sections~\ref{sec:movielens} and~\ref{sec:ebay}).


\subsection{Segmentation algorithm for general preference graph $\Pcal$} 
\label{sec:gen_graph}
We now focus on the case when the number of observations may be different across customers. The key issue is that the log-likelihood values $-\log \mi{f}_{\pool}(\xiobs)$ depend on the number of observations $d_i$ and therefore, should be appropriately normalized in order to be meaningfully compared across customers. A natural way is to normalize the log-likelihood value of customer $i$ by the corresponding degree $d_i$, which results in Algorithm~\ref{alg:lreg_seg} but applied to the unequal degree setting. Such degree normalization is appropriate when the observations across items are independent, so that the pooled distribution $f_{\pool}(\bmx)$ has a product form $f_{\pool, 1}(x_1) \cdot f_{\pool, 2}(x_2) \dotsb f_{\pool, n}(x_n)$. In this case, the log-likelihood under the pooled model becomes $\log \mi{f}_{\pool}(\xiobs) = \sum_{j \in N(i)} \log f_{\pool,j}(x_{ij})$, which scales in the number of observations $d_i$.

The degree normalization, however, does not take into account any dependence structure in the item labels. For instance, consider the extreme case when the observations across all items are perfectly correlated under the pooled model, such that customers either like all items or dislike all items with probability 0.5 each. In this case, the log-likelihood does not depend on the number of observations, but the degree normalization unfairly penalizes customers with few observations.
To address this issue, we use entropy-based normalization:
\begin{equation}
\label{eq:proj_def}
    \proj_i = \frac{-\log \mi{f}_{\pool}(\xiobs)}{H(\mi{f}_{\pool})} = \frac{-\log \mi{f}_{\pool}(\bmx_i^{\obs})}{-\sum_{\bmy \in \pci} \mi{f}_{\pool}(\bmy) \log \mi{f}_{\pool}(\bmy)}
\end{equation}
where $H(\mi{f}_{\pool})$ denotes the {\em entropy} of distribution $\mi{f}_{\pool}$. When the observations across items are i.i.d.\ it can be seen that entropy-based normalization reduces to degree-normalization, upto constant factors. The key benefit of the entropy normalization is that when the population is homogeneous (i.e.\ consists of a single segment), it can be shown that the projection scores of all customers concentrate around $1$ (see the discussion in Section~\ref{sec:lc_ind}). Consequently, deviations in the projection scores from $1$ indicate heterogeneity in the customer population and allows us to identify the different~segments.

Entropy-based normalizations have been commonly used in the literature for normalizing mutual information~\citep{strehl2002cluster}---our normalization is inspired by that. In addition to accounting for dependency structures within the pooled distribution, it has the effect of weighting each observation by the strength of the evidence it provides. Specifically, because the log-likelihood value $-\log \mi{f}_{\pool}(\bmx_i^{\obs})$ only provides incomplete evidence of how well $f_\pool$ captures the preferences of customer $i$ when there are missing observations, we assess the confidence in the evidence by dividing the log-likelihood value by the corresponding entropy $H(\mi{f}_{\pool})$ of the distribution $\mi{f}_{\pool}$. Higher values of entropy imply lower confidence. Therefore, when the entropy is high, the projection score will be low, indicating that we don't have sufficient evidence that customer $i$'s observations are not well-explained by $f_\pool$.
The algorithm with entropy-based normalization is summarized in Algorithm~\ref{alg:gen_seg}.

\begin{algorithm}
\caption{Segmentation algorithm with entropy normalization}
\label{alg:gen_seg}
\begin{algorithmic}
\STATE {\em same as Algorithm~\ref{alg:lreg_seg} except replace step 3 with}: \\
Compute projection score of customer $i$ using equation~\eqref{eq:proj_def}
\end{algorithmic}
\end{algorithm}
\vspace{-1em}
We note that the entropy may be difficult to compute because, in general, it requires summing over an exponentially large space. For such cases, either the entropy may be computed approximately using existing techniques in probabilistic graphical models~\citep{hinton2002training,salakhutdinov2007restricted} or the degree normalization of Algorithm~\ref{alg:lreg_seg} may be used as an approximation.

\subsection{Segmentation algorithm for partially specified model families}
\label{sec:item_cat}

The discussion so far assumed that the model family $\F$ is fully-specified, i.e., for a given value of $\omega \in \Omega$, the model $f(\cdot;\omega)$ completely specifies how an observation vector $\bm{x}$ is generated. In practice, such complete specification may be especially difficult to provide when the item universe is large (for instance, millions in our eBay case study) and there are complex cross-effects among items, such as the correlation between the rating and purchase signal for the same item or complementarity effects among products from related categories (clothing, shoes, accessories, etc.). 
To address this issue, we extend our algorithm to the case when the model is only partially specified.


We assume that the universe $[n]$ of items is partitioned into $B > 1$ ``categories'' $\set{\mathcal{I}_1, \mathcal{I}_2, \dotsc, \mathcal{I}_B}$ such that $\mathcal{I}_b$ is the set of items in category $b \in [B] \defas \set{1,2,\ldots,B}$ and containing $n_b$ items. A model describing the observations within each category is specified, but any interactions across categories are left unspecified. We let $\F_b(\Omega_b) = \set{f(\bmx_b; \omega) \; \colon \; \omega \in \Omega_b}$ denote the model class for category $b$, so that segment $k$ is characterized by the $B$ models $(f_{k1}, f_{k2}, \ldots, f_{kB})$ with $f_{kb} \in \F_b$ for all $1 \leq b \leq B$. Further, $\xobs_{ib}$ denotes the vector of observations of customer $i$ for items in category $b$; if there are no observations, we set $\xobs_{ib} = \phi$.

\newcommand*{\horzbar}{\rule[.5ex]{2.5ex}{0.5pt}}

Under this setup, we run our segmentation algorithm (Algorithm~\ref{alg:lreg_seg} or \ref{alg:gen_seg}) separately for each category of items. This results in a $B$-dimensional representation $\bm{\proj}_i = (\proj_{i1}, \proj_{i2}, \ldots, \proj_{iB})$ for each customer $i$, where $\proj_{ib}$ is the representation computed by our algorithm for customer $i$ and category $b$. When $\xobs_{ib} = \phi$, we set $\proj_{ib} = \phi$. 
We represent these vectors compactly as the following $m \times B$ matrix with row $i$ corresponding to~customer~$i$:
\[ 
\projmat = \begin{bmatrix}
    \proj_{11} & \proj_{12} & \dots  & \proj_{1B} \\
    \proj_{21} & \proj_{22}  & \dots  & \proj_{2B} \\
    \vdots & \vdots & \ddots & \vdots \\
    \proj_{m1} & \proj_{m2} & \dots  & \proj_{mB}
    \end{bmatrix} \]
When matrix $\projmat$ is complete, the algorithm stops and outputs $\projmat$. When it is incomplete, we obtain low-dimensional representations for the customers using low-rank matrix decomposition (or factorization) techniques, similar to those adopted in collaborative filtering applications~\citep{rennie2005fast,mnih2007probabilistic,koren2009matrix}. These techniques assume that the matrix $\projmat$ with missing entries can be factorized into a product of two low-rank matrices---one specifying the customer representation and the other an item representation, in a low dimensional space. The low-rank structure naturally arises from assuming that only a small number of (latent) factors influence the cross-effects across the categories. With this assumption, we compute an $r$-dimensional representation $\bm{u}_i \in \Real^{r}$ for each customer $i$ and $\bm{v}_b \in \Real^r$ for each item category $b$ by solving the following optimization problem:
\begin{equation}
\label{eq:rankr_fact}
\min_{\bmU, \bmV} \sum \limits_{i=1}^m \sum \limits_{b=1}^B \ind[\proj_{ib} \neq \phi] \left(\proj_{ib} - \bm{u}_i^T\bm{v}_b \right)^2
\end{equation}
where matrices $\bm{U} \in \Real^{m \times r}$ and $\bm{V} \in \Real^{B \times r}$ are such that
\[ 
\bmU = \begin{bmatrix}
    \horzbar & \bm{u}_1^T & \horzbar\\
    \horzbar & \bm{u}_2^T & \horzbar \\
     & \vdots &  \\
    \horzbar & \bm{u}_m^T & \horzbar 
        \end{bmatrix} 
\;\;\;\;\; \;\;\;\;\        
    \bmV = \begin{bmatrix}
    \horzbar & \bm{v}_1^T & \horzbar\\
    \horzbar & \bm{v}_2^T & \horzbar \\
     & \vdots &  \\
    \horzbar & \bm{v}_B^T & \horzbar 
        \end{bmatrix}       
    \]
Note that the rank $r \ll \min(m, B)$. When the number of customers $m$ or categories $B$ is large, computing the low-rank decomposition may be difficult. But there has been a lot of recent work to develop scalable techniques for such matrices (as in collaborative filtering applications like Netflix), see the works of~\cite{takacs2009scalable, mazumder2010spectral} and references therein. The precise algorithm is summarized in Algorithm~\ref{alg:cat_full_seg}.

\begin{algorithm}
\caption{Segmentation algorithm when the model class $\F$ is partially specified}
\label{alg:cat_full_seg}
\begin{algorithmic}[1]
\STATE \textbf{Input:} observed labels $\bm{x}_1^{\obs}, \bm{x}_2^{\obs},\ldots, \bm{x}_m^{\obs}$, item partitioning $\set{\mathcal{I}_1, \dotsc, \mathcal{I}_B}$, model family $\F_b$ for each $1 \leq b \leq B$, the rank $r \ll \min (m, B)$ of low-rank representation
\STATE $f_{\pool,b} \gets$ estimated pooled model from family $\F_b$ for all $1 \leq b \leq B$
\STATE Compute $\proj_{ib}$ for each customer $i$ and category $b$ using Algorithm~\ref{alg:lreg_seg} or~\ref{alg:gen_seg} whenever $\xobs_{ib} \neq \phi$ otherwise set $\proj_{ib} = \phi$
\STATE Create the $m \times B$ matrix $\projmat$ where row $i$ represents the projection score vector of customer $i$, $\bm{\proj}_i = (\proj_{i1}, \proj_{i2}, \ldots, \proj_{iB})$
\STATE If $\projmat$ is incomplete, compute rank $r$-factorization $\projmat \approx \bm{U}\bm{V}^T$ where $\bm{U} \in \Real^{m \times r}, \bm{V} \in \Real^{B \times r}$ by solving the optimization problem in~\eqref{eq:rankr_fact}.
\STATE \textbf{Output:} $\projmat$ if it is complete and $\bmU\bmV^T$ otherwise
\end{algorithmic}
\end{algorithm}

We conclude this section with a few remarks on how to obtain the segments from the representations of customers obtained by our algorithms. Let $R$ denote the dimension of the representation of each customer. It follows from our descriptions above that $R = 1$ when the model class is fully specified and $R = B$ when the model class is only partially specified. When $R \ll n$, the $k$-means algorithm or spectral clustering~\citep{von2007tutorial} with Euclidean distances may be used to cluster the representations of the customers. In Section~\ref{sec:results}, we show that these techniques successfully recover the underlying segments under standard assumptions. When $R$ is large, the curse of dimensionality kicks in and traditional distance-based clustering algorithms may not be able to meaningfully differentiate between similar and dissimilar projection score vectors (see~\citealp{aggarwal2001surprising} and~\citealp{steinbach2004challenges}). For this setting, we recommend using the spectral projection technique, which projects the $m \times R$ matrix to the subspace spanned by its top $K$ principal components and then clusters the projections in the lower-dimensional space. This technique was proposed in the theoretical computer science literature~\citep{vempala2004spectral,achlioptas2005spectral,kannan2005spectral} to cluster observations generated from a mixture of multivariate Gaussians. The projections can be shown to preserve the separation between the mean vectors and at the same time ensure (tighter) concentration of the original samples around the respective means, resulting in more accurate segmentation.

\section{Theoretical Results}
\label{sec:results}

\newcommand{\Ber}{{\rm Ber}}
\newcommand{\ztheta}{\hbmz_{\theta_0,\theta_1, \ldots, \theta_K}}
\newcommand{\rproj}{\bm{\mathrm{PSCORE}}}

Our segmentation algorithm is analytically tractable and in this section, we derive theoretical conditions for how ``separated'' the underlying segments must be to guarantee asymptotic recovery using our algorithm. Our results are similar in spirit to existing theoretical results for clustering observations from mixture models, such as mixture of multivariate Gaussians~\citep{kannan2005spectral}.

For the purposes of the theoretical analysis, we focus on the following standard setting---there is a population of $m$ customers comprising $K$ distinct segments such that a proportion $q_k$ of the population belongs to segment $k$, for each $k \in [K]=\set{1,2,\ldots,K}$. Segment $k$ is described by distribution $\pi_k \colon \bin^n \to [0, 1]$ over the domain $\mathcal{B} := \bin^n$. Note that this corresponds to the scenario when $\mathcal{X}_j = \bin$ for all items $j$ (see the notation in Section~\ref{sec:setup}). We frequently refer to $+1$ as {\em like} and $-1$ as {\em dislike} in the remainder of the section.
Customer $i$'s latent segment is denoted by $z_i \in [K]$, so that if $z_i = k$, then $i$
samples a vector $\bm{x}_i \in \mathcal{B}$ according to distribution $\pi_k$, and then assigns the label $x_{ij}$ for item $j$. 
We focus on asymptotic recovery of the true segment labels $\bm{z} = (z_1, z_2, \ldots,z_m)$, as the number of items $n \to \infty$.

The performance of our algorithm depends on the separation among the hyper-parameters describing the segment distributions $\pi_k$, as well as the number of data points available per customer. Therefore, we assume that the segment distributions are ``well-separated'' (the precise technical conditions are described below) and the number of data points per customer goes to infinity as $n \to \infty$.
The proofs of all statements are in the Appendix.

\subsection{Fully specified model family: independent item preferences}
\label{sec:lc_ind}
We first consider the case where $\pi_k$ belongs to a fully specified model family $\F(\Omega)$, such that customer labels across items are independent.
More precisely, we have the following model:

\begin{definition}[Latent Class Independent Model ({\sf LC-IND})] \label{def:lc_ind}
Each segment $k$ is described by distribution $\pi_k \colon \bin^n \to [0, 1]$ such that labels $\set{x_{j}}_{j \in [n]}$ are independent and identically distributed. Denote $\alpha_k = \Pr_{\bmx \sim \pi_k}[x_{j} = +1]$ for all items $j \in [n]$, i.e. $\alpha_k$ is the probability that a customer from segment $k$ likes an item. Customer $i$ in segment $k$ samples vector $\tilde{\bm{x}}_{i}$ according to distribution $\pi_{k}$ and provides label $\tilde{x}_{ij}$ for item $j$.
\end{definition}

We assume that the segment parameters are bounded away from 0 and 1, i.e. there exists a constant $\alphab_{\min} > 0$ such that $0 < \alphab_{\min} \leq \alpha_k \leq 1 - \alphab_{\min} < 1$ for all segments $k \in [K]$. 
Further, let $H(\beta_1, \beta_2) = -\beta_1 \log \beta_2 - (1-\beta_1) \log (1-\beta_2)$ denote the {\em cross-entropy} between the Bernoulli distributions $\Ber(\beta_1)$ and $\Ber(\beta_2)$ and $H(\alpha) = -\alpha \log(\alpha) - (1-\alpha) \log(1-\alpha)$ denote the binary entropy function, where $0 \leq \alpha \leq 1$. Let $\rproj_i$ denote the (uni-dimensional) projection score computed by Algorithm~\ref{alg:gen_seg}, note that it is a random variable under the above generative model. 

Given the above, we derive necessary and sufficient conditions to guarantee (asymptotic) recovery of the true customer segments under the {\sf LC-IND} model. 

\subsubsection{Necessary conditions for recovery of true segments.} We first prove an important lemma concerning the concentration of the customer projection scores computed by our algorithm.

\begin{lemma}[Concentration of customer projection scores under {\sf LC-IND} model]
Given a customer population $[m]$ and collection of items $[n]$, suppose that the preference graph $\Pcal$ is $\ell$-regular, i.e. $d_i = \ell$ for all customers $1 \leq i \leq m$.
Define the parameter $\alphapool \defas \sum_{k=1}^K q_k \alpha_k$. 
Then given any $0 < \varepsilon < 1$, the projection scores computed by Algorithm~\ref{alg:gen_seg} are such that:
{\footnotesize
\begin{equation*}
\Pr\expt{\abs{\rproj_i - \frac{H(\alphab_{z_i}, \alpha_{\rm pool})}{H(\alpha_{\rm pool})}} > \varepsilon \frac{H(\alphab_{z_i}, \alpha_{\rm pool})}{H(\alpha_{\rm pool})}} \leq 4\exp\left(\frac{-2 \ell \varepsilon^2 \alphab_{\min}^2}{81}\right)
+ 12\exp\left( \frac{-2 m\cdot \ell \cdot  \varepsilon^2 \alphabpool^2 \log^2 \left(1- \alphabpool \right))}{81 \big(1-\log \left(1- \alphabpool \right)\big)^2}\right)
\end{equation*}}
where $\alphabpool \defas \min \left(\alphapool, 1-\alphapool\right)$. In other words, the projection scores of customers in segment $k$ concentrate around the ratio $\frac{H(\alphab_k, \alpha_{\rm pool})}{H(\alpha_{\rm pool})}$, with high probability as the number of observations from each customer $\ell \to \infty$.
\label{thm:lc_ind_conc}
\end{lemma}

Lemma~\ref{thm:lc_ind_conc} reveals the necessary conditions our algorithm requires to recover the true customer segments. To understand the result, first suppose that $\alpha_{\rm pool} \neq (1/2)$. Then, the above lemma states that the model-based projection scores of customers in segment $k$ concentrate around $\frac{H(\alpha_k, \alpha_{\rm pool})}{H(\alpha_{\rm pool})}$ which is proportional to $-\alpha_k\log\frac{\alpha_{\rm pool}}{1-\alpha_{\rm pool}} - \log(1-\alpha_{\rm pool})$. Consequently, we require that $\alpha_{k} \neq \alpha_{k'}$ whenever $k \neq k'$ to ensure that the projection scores of customers in different segments concentrate around distinct values.
%
The result also states that the projection scores of customers with similar preferences (i.e. belonging to the same segment) are close to each other, i.e. concentrate around the same quantity, whereas the scores of customers with dissimilar preferences (i.e. belonging to different segments) are distinct from each other. For this reason, although it is not a priori clear, our segmentation algorithm is consistent with the classical notion of distance- or similarity-based clustering, which attempts to maximize intra-cluster similarity and inter-cluster dissimilarity. When $\alpha_{\rm pool} = (1/2)$, it follows that $H(\alpha_k, \alphapool) = H(\alphapool)$ for {\em any} $0 \leq \alpha_k \leq 1$, and therefore all the customer projection scores concentrate around $1$. In this scenario, our algorithm cannot separate the customers even when the parameters $\alpha_k$ of different segments are distinct. Note that $\alpha_{\rm pool} = \sum_k q_k \alpha_k$, which is the probability that a random customer from the population likes an item. Therefore, when $\alpha_{\rm pool} = (1/2)$, the population is indifferent in its preferences for the items, resulting in all the customers being equidistant from the pooled customer. 

The above discussion leads to the following theorem.

\begin{theorem}[Necessary conditions for recovery of true segments under {\sf LC-IND} model]
Under the setup of Lemma 1, the following conditions are necessary for recovery of the true customer segments:

$1$. All segment parameters are distinct, i.e. $\alpha_k \neq \alpha_{k'}$ whenever $k \neq k'$, and $2$. $\alphapool \neq \frac{1}{2}$.
\end{theorem}

It is easy to see that the first condition is necessary for {\em any} segmentation algorithm. We argue that the second condition, i.e. $\alphapool \neq \frac{1}{2}$, is also necessary for the standard latent class segmentation technique. Specifically, consider two segments such that $q_1 = q_2 = 0.5$ and let $\alpha_1 = 1, \alpha_2 = 0$. Then, it follows that $\alphapool = q_1 \alpha_1 + q_2 \alpha_2 = \frac{1}{2} \cdot 1 + \frac{1}{2} \cdot 0 = \frac{1}{2}$. Let us consider only a single item, i.e. $n = 1$. Then, under this parameter setting, all customers in segment $1$ will give the label $+1$ and all customers in segment $2$ will give label $-1$. Recall that the latent class method estimates the model parameters by maximizing the log-likelihood of the observed labels, which in this case looks like:
\[ \log \mathcal{L} = \frac{m}{2} \log \left(q_1 \alpha_1 + q_2 \alpha_2\right) + \frac{m}{2} \log \left(q_1 \cdot (1-\alpha_1) + q_2 \cdot (1-\alpha_2)\right) \]
Then it can be seen that the solution $\hat{q}_1 = \hat{q}_2 = 0.5$ and $\hat{\alpha}_1 = \hat{\alpha}_2 = 0.5$ achieves the optimal value of the above log-likelihood function, and therefore is a possible outcome recovered by the latent class method. This shows that the condition $\alphapool \neq \frac{1}{2}$ is also necessary for the standard latent class segmentation technique.

We note that our results readily extend to the case when $\Pcal$ is not $\ell$-regular but with additional~notation.

\subsubsection{Sufficient conditions for recovery of true segments.}
Having established the necessary conditions, we now discuss the asymptotic {\em misclassification rate}, defined as the expected fraction of customers incorrectly classified, of our algorithm. In order to analyze the misclassification rate, we consider the following nearest-neighbor (NN) classifier $\hbmz(\cdot)$, where customer $i$ is classified as:
	\begin{equation*}
		\hbmz(i) = \argmin_{k = 1,2,\ldots,K} \frac{\abs{\rproj_i - H_k}}{H_k}
	\end{equation*}
	where $H_k \defas \frac{H(\alpha_k, \alphapool)}{H(\alphapool)}$. Note that $H_k > 0$ since $0 < \alpha_{\min} \leq \alpha_k \leq 1- \alpha_{\min} < 1$, for all $k \in [K]$.



Given the necessary conditions established above and to ensure that we can uniquely identify the different segments, we assume that the segments are indexed such that $\alpha_1 < \alpha_2 < \ldots < \alpha_K$.
Then, we can prove the following recoverability result:

\begin{theorem}[Asymptotic recovery of true segments under {\sf LC-IND} model]
\label{thm:asymp_rec}
Under the setup of Lemma 1, suppose $0 < \alpha_{\min} \leq \alpha_1 < \alpha_2 < \dotsb < \alpha_K \leq 1 - \alpha_{\min}$ and $\alphapool \neq \frac{1}{2}$. Further, denote 
$\lambda = \min_{k=1,2,\ldots,K-1} (\alpha_{k+1} - \alpha_k)$. 
Given any $0 < \delta < 1$, suppose that
\[ \ell \geq \frac{648}{\lambda^2} \cdot \left(\frac{\log \alpha_{\min}}{\log (1-\amin) \cdot \amin}\right)^2 \cdot \frac{1}{\log^2\frac{\alphapool}{1-\alphapool}} \cdot \log (16/\delta) \]
Then, it follows that 
\[ \frac{1}{m} \sum \limits_{i=1}^m \Pr\expt{\hbmz(i) \neq z_i} < \delta \]
Further, when $\ell = \log n$ and $m \geq \left(\frac{1 - \log (1-\alphabpool)}{\log(1 - \alphabpool)}\right)^2$, we have:
	\begin{align*}
		\frac{1}{m} \sum \limits_{i=1}^m \Pr\expt{\hbmz(i) \neq z_i} = O\left(n^{-\frac{2\Lambda^2 \alphab_{\min}^2}{81} }\right) \;\;\; \text{where the constant }
		 \Lambda \defas \frac{\lambda}{2} \cdot \left(\frac{\abs{\log \frac{\alphapool}{1-\alphapool}}}{\abs{\log \amin}}\right)
		\end{align*}
\end{theorem}
Theorem~\ref{thm:asymp_rec} provides an upper bound on the misclassification rate of our segmentation algorithm in recovering the true customer segments. The first observation is that as the number of labels from each customer $\ell \to \infty$, the misclassification rate of the NN classifier goes to zero. The result also allows us determine the number of samples $\ell$ needed per customer to guarantee an error rate $\delta$. In particular, $\ell$ depends on three quantities:
\begin{enumerate}
    \item $\frac{1}{\lambda^2}$ where $\lambda$ is the minimum separation between the segment parameters. This is intuitive---the ``closer'' the segments are to each other (i.e. smaller value of $\lambda$), the more samples are required per customer to successfully identify the true segments.
    \item $\frac{1}{\log^2\frac{\alphapool}{1-\alphapool}}$ where recall that $\alphapool$ is the probability that a random customer from the population likes an item. If $\alphapool \approx \frac{1}{2}$, then $\log^2\frac{\alphapool}{1-\alphapool} \approx 0$ so that we require a large number of samples per customer. As $\alphapool$ deviates from $\frac{1}{2}$, the quantity $\log^2\frac{\alphapool}{1-\alphapool}$ increases, so fewer samples are sufficient. This also makes sense---when $\alphapool = (1/2)$, our algorithm cannot identify the underlying segments, and the farther $\alphapool$ is from $(1/2)$, the easier it is to recover the true segments.
    \item $\amin$, where as $\amin \to 0$, the number of samples required diverges. Note that $\amin$ (resp. $1-\amin$) specifies a lower (upper) bound on the segment parameters $\alpha_k$---a small value of $\amin$ indicates that there exists segments with values of $\alpha_k$ close to either $0$ or $1$; and since the number of samples required to reliably estimate $\alpha_k$ (resp. $1-\alpha_k$) grows as $\frac{1}{\alpha_k^2}$ (resp. $\frac{1}{(1-\alpha_k)^2}$), $\ell$ must diverge as  $\amin \to 0$.
\end{enumerate}

Our result shows that as long as each customer provides atleast $\log n$ labels, the misclassification rate goes to zero, i.e. we can accurately recover the true segments with high probability, as the number of items $n \to \infty$. Although the number of labels required from each customer must go to infinity, it must only grow logarithmically in the number of items $n$. Further, this holds for {\em any} population size ``large enough''.

Note that the NN classifier above assumes access to the ``true'' normalized cross-entropies $H_k$. In practice, we use ``empirical'' NN classifiers, which replace $H_k$ by the corresponding cluster centers of the projection scores. Lemma~\ref{thm:lc_ind_conc} guarantees the correctness of this approach under appropriate assumptions, because the projection scores of segment $k$ customers concentrate around $H_k$.

\subsection{Partially specified model family: independent within-category item preferences}
\label{sec:lc_ind_cat}
We can extend the results derived above to the case when the distributions $\pi_k$ belong to a partially specified model family, as discussed in Section~\ref{sec:item_cat}. Specifically, suppose that the item set $[n]$ is partitioned into $B > 1$ (disjoint) categories: $\I_1 \cupdot \I_2 \dotsb \cupdot \I_B$. The preferences of customers vary across the different categories, specifically we consider the following generative model:

\begin{definition}[Latent Class Independent Category Model ({\sf LC-IND-CAT})]\label{def:lc_ind_cat}
Each segment $k$ is described using distribution $\pi_{k} \colon \bin^n \to [0, 1]$ such that labels $\set{x_{j_b}}_{j_b\in \I_b}$ for items within a single category $b \in [B]$ are independent and identically distributed; but labels for items in different categories can have arbitrary correlations. Let $\bm{\alpha}_{k} = (\alpha_{k1}, \alpha_{k2}, \dotsb, \alpha_{kB})$ be such that $\Pr_{\bmx \sim \pi_k}[x_{j_b} = +1] = \alpha_{kb}$ for each item $j_b \in \I_b$ and each category $b \in [B]$. Customer $i$ in segment $k$ samples vector $\tilde{\bm{x}}_{i}$ according to distribution $\pi_{k}$ and provides label $\tilde{x}_{ij}$ for each item $j$.
\end{definition}

The above model is general and can be used to account for correlated item preferences (as opposed to independent preferences considered in section~\ref{sec:lc_ind}). As a specific example, suppose that for each item, we have two customer observations available: whether the item was purchased or not, and a like/dislike rating (note that one of these can be missing). Clearly these two observations are correlated and we can capture this scenario in the {\sf LC-IND-CAT} model as follows: there are two item ``categories''---one representing the purchases and the other representing the ratings. In other words, we create two copies of each item and place one copy in each category. Then, we can specify a joint model over the item copies such that
purchase decisions for different items are independent, like/dislike ratings for different items are also independent but the purchase decision and like/dislike rating for the {\em same} item are dependent on each other.
Similar transformations can be performed if we have more observations per item or preferences are correlated for a group of items. Therefore, the above generative model is fairly broad and captures a wide variety of customer preference structures.

As done for {\sf LC-IND} model, we assume that the underlying segment parameters are bounded away from 0 and 1, i.e. there exists constant $\alphab_{\min} > 0$ such that $0 < \alphab_{\min} \leq \alphab_{kb} \leq 1 - \alphab_{\min} < 1$ for all segments $k \in [K]$, and all item categories $b \in [B]$. 
Let $d_{ib} > 0$ be the number of observations for customer $i$ in category $b$ and 
let $\projrvec_i$ denote the projection score vector computed by Algorithm~\ref{alg:cat_full_seg} for customer $i$, note that it is a $B$-dimensional random vector under the generative model above. 

\subsubsection{Necessary conditions for recovery of true segments.} We first state an analogous concentration result for the customer projection score vectors computed by our algorithm.
\begin{lemma}[Concentration of projection score vectors under {\sf LC-IND-CAT} model]
For a population of $m$ customers and $n$ items partitioned into $B > 1$ categories, suppose that the preference graph $\Pcal$ is such that each customer labels exactly $\ell_b > 0$ items in category $b$, i.e. $d_{ib} = \ell_b$ for all $1 \leq i \leq m$.
Define the parameters $\alphacatpool{b} \defas \sum_{k=1}^K q_{k} \alpha_{kb}$ for each item category $b$, $\ell_{\min} \defas \min_{b \in [B]} \ell_b$, and $\hat{\alpha}_{\rm pool}\defas \min_{b \in [B]} \bar{\alpha}_{b, {\rm pool}}$ where $\bar{\alpha}_{b, {\rm pool}} \defas \min \left(\alphacatpool{b}, 1-\alphacatpool{b} \right)$. Then given any $0 < \varepsilon < 1$, the projection score vectors computed by Algorithm~\ref{alg:cat_full_seg} are such that:
{\footnotesize
\begin{equation*}
\Pr\expt{\norm{\projrvec_i - \bm{H}_{z_i}}_1 > \varepsilon \norm{\bm{H}_{z_i}}_1} \leq 4\cdot B \cdot \exp\left(\frac{-2 \ell_{\min} \varepsilon^2 \alphab_{\min}^2}{81}\right)
+ 12 \cdot B \cdot \exp\left( \frac{-2 m\cdot \ell_{\min} \cdot \varepsilon^2 \hatalphapool^2 \log^2 (1-\hatalphapool)}{81 \big(1- \log (1-\hatalphapool)\big)^2}\right)
\end{equation*}}
\label{thm:lc_ind_cat_conc}
\end{lemma}
In the lemma above, $\bm{H}_k = (H_{k1}, H_{k2}, \dotsb, H_{kB})$ is a $B$-dimensional vector such that $H_{kb} = \frac{H(\alphab_{kb}, \alphacatpool{b})}{H(\alphacatpool{b})}$ (recall notation from section~\ref{sec:lc_ind}) and $\norm{\cdot}_1$ denotes the $\mathcal{L}_1$-norm. Lemma~\ref{thm:lc_ind_cat_conc} implies the following necessary conditions:
\begin{theorem}[Necessary conditions for recovery of true segments under {\sf LC-IND-CAT} model]
\label{thm:nec_lc_ind_cat}
Under the setup of Lemma 2, the following conditions are necessary for recovery of the true customer segments:
\begin{enumerate}
\item $\alphacatpool{b} \neq \frac{1}{2}$ for some category $b \in [B]$.
\item Let $B' = \set{b \in [B] \; \colon \; \alphacatpool{b} \neq \frac{1}{2}}$ and denote $\left(\alphavec_{k}\right)_{b \in B'}$ as the sub-vector consisting of components corresponding to item categories $B'$. 
Then $\left(\alphavec_k\right)_{b \in B'} \neq \left(\alphavec_{k'}\right)_{b \in B'}$ whenever $k \neq k'$.
\end{enumerate}
\end{theorem}
Similar to the {\sf LC-IND} case, $\alphacatpool{b} = (1/2)$ for {\em all} item categories implies that the population is indifferent over items in all the categories. However, we require the population to have well-defined preferences for at least one category in order to be able to separate the segments.
Further,
since $H_{kb} \propto -\alphab_{kb} \log \frac{\alphacatpool{b}}{1 - \alphacatpool{b}} - \log (1- \alphacatpool{b})$, we need $\alphab_{kb} \neq \alphab_{k'b}$ for at least one item category $b$ where $\alphacatpool{b} \neq \frac{1}{2}$ to ensure that the vectors $\bm{H}_k$ and $\bm{H}_{k'}$ are~distinct, for any two segments $k \neq k'$. 

\subsubsection{Sufficient conditions for recovery of true segments.}
As for the case of the {\sf LC-IND} model, we consider another NN classifier to evaluate the asymptotic misclassification rate of our segmentation algorithm, where customer $i$ is classified as:
	\begin{equation*}
		\hbmz_2(i) = \argmin_{k = 1,2,\ldots,K} \frac{\norm{\projrvec_i - \bm{H}_k}_1}{\norm{\bm{H}_k}_1}
	\end{equation*}

Given the above necessary conditions, we can prove the following recoverability result:

\begin{theorem}[Asymptotic recovery of true segments under {\sf LC-IND-CAT} model]
\label{thm:asymp_rec_cat}
Suppose that the conditions in Theorem 3 are satisfied. Denote $\bm{w} = (w_1, w_2, \dotsb, w_B)$ with $w_b = \abs{\log \frac{\alphacatpool{b}}{1-\alphacatpool{b}}}$ and
$\gamma = \min_{k \neq k'} \norm{\bm{w} \odot (\bm{\alpha}_{k} - \bm{\alpha}_{k'})}_1$ where $\odot$ represents element-wise product. 
Under the setup of Lemma 2 and given any $0 < \delta < 1$, suppose that
\[ 
\ell_{\min} \geq \frac{648 B^2}{\gamma^2} \cdot \left(\frac{\log \alpha_{\min}}{\log^2 (1-\amin) \cdot \amin}\right)^2 \log (16B/\delta) \]
Then, it follows that 
\[ \frac{1}{m} \sum \limits_{i=1}^m \Pr\expt{\hbmz_2(i) \neq z_i} < \delta \]
Further, when $\ell_{\min} = \log n$ and $m \geq  \left(\frac{1 - \log (1-\hatalphapool)}{\log(1 - \hatalphapool)}\right)^2$, for fixed $B$ we have:
	\begin{align*}
		\frac{1}{m} \sum \limits_{i=1}^m \Pr\expt{\hbmz_2(i) \neq z_i} = O\left(n^{\frac{-2\Gamma^2 \alphab_{\min}^2}{81}}\right)  
		\;\;\; \text{where the constant } \Gamma \defas \frac{\gamma}{2B} \cdot \abs{\frac{\log (1-\amin)}{\log \alpha_{\min}}}
		\end{align*}
\end{theorem}

We make a few remarks about Theorem~\ref{thm:asymp_rec_cat}. First, as $\ell_{\min} \to \infty$, i.e. the number of labels in each item category $\ell_b \to \infty$, the misclassification rate of the NN classifier goes to zero. Second, to achieve misclassification rate of atmost $\delta$, the number of samples $\ell_{\min}$ scales as
\begin{enumerate}
    \item $\frac{1}{\gamma^2}$ where $\gamma$ is the minimum weighted $\mathcal{L}_1$-norm of the difference between the parameter vectors of any two segments. This is similar to a standard ``separation condition''---the underlying segment vectors $\alphavec_k$ should be sufficiently distinct from each other, as measured by the $\mathcal{L}_1$-norm. However, instead of the standard $\mathcal{L}_1$-norm, we require a weighted form of the norm, where the weight of each component is given by $w_b = \abs{\log \frac{\alphacatpool{b}}{1 - \alphacatpool{b}}}$. If $\alphacatpool{b} \approx \frac{1}{2}$, then $w_b \approx 0$ so that the separation in dimension $b$ is weighed lower than categories where $\alphacatpool{b}$ is sufficiently distinct from $\frac{1}{2}$. This follows from the necessary condition in Theorem~\ref{thm:nec_lc_ind_cat} and is a consequence of the simplicity of our algorithm that relies on measuring deviations of customers from the population preference.
    \item $B^2$ where $B$ is the number of item categories. This is expected---as the number of categories increases, we require more samples to achieve concentration in {\em all} the dimensions of the projection score vector $\projrvec_i$.
    \item $\amin$, the dependence on which is similar to the {\sf LC-IND} model case, but with an extra factor of $\log^2 (1-\amin)$ in the denominator, indicating a more stronger dependence on $\amin$.
\end{enumerate}
 Finally, it follows that a logarithmic number of labels in {\em each} category is sufficient to guarantee recovery of the true segments with high probability as the total number of items $n \to \infty$, provided the population size $m$ is ``large enough''.

\section{Computational study: Accuracy of model-based projection technique}
\label{sec:simul_exp}
\newcommand{\mcOne}[1]{\multicolumn{1}{c}{#1}}

In this section, we use synthetic data to analyze the misclassification rate of our segmentation algorithm as a function of the number of labels available per customer. We compare our approach to the standard latent class (LC) method, which uses the EM algorithm to estimate posterior segment membership probabilities of the different customers (we discuss the method in more detail below). The results from the study demonstrate that our approach (1) outperforms the LC benchmark by upto $28\%$ in recovering the true customer segments and is more robust to high levels of sparsity in the customer labels and (2) is fast, with upto $11\times$ gain in computation time compared to the LC method.


{\bf Setup}. We chose $m=2000$ customers and $n=100$ items and considered the following standard latent class generative model: There are $K$ customer segments with $q_k$ denoting the proportion of customers in segment $k$; we have $q_k > 0$ for all $1 \leq k \leq K$ and $\sum_{k=1}^K q_k = 1$. The customer-item preference graph follows the standard Erd\H{o}s-R\'enyi (Gilbert) model with parameter $0 < p < 1$: each edge $(i,j)$ between customer $i$ and item $j$ is added independently with probability $p$. The parameter $1 - p$ quantifies the {\em sparsity} of the graph: higher the value of $1-p$, sparser the graph. All customers in segment $k \in [K]$ generate binary labels as follows: given parameter $\alpha_k \in (0,1)$, they provide rating $+1$ to item $j$ with probability $\alpha_k$ and rating $-1$ with probability $1- \alpha_k$.

We denote each ground-truth model type by the tuple: $(K, 1- p)$. We generated 15 models by varying $K$ over the set $\set{5, 7, 9}$ and $1-p$ over the set $\set{0, 0.2, 0.4, 0.6, 0.8}$. For each value of $K$, we sample the segment proportions from a Dirichlet distribution with parameters $\beta_1 = \beta_2 = \dotsb = \beta_K = K+1$ which tries to ensure that all segments have sufficiently large sizes by placing a high mass on equal proportions. Similarly for each $K$, the parameters $\alpha_k$ are chosen as $\alpha_k = 0.05 + 0.9(k-1)/(K-1)$ ($K$ uniformly spaced points in the interval $[0.05, 0.95]$) for all $1 \leq k \leq K$.

For each ground-truth model type, we randomly generated 30 model instances as follows: (a) randomly partition the customer population into $K$ segments with segment $k$ having proportion $q_k$; (b) randomly generate the customer-item preference graph by adding edge $(i,j)$ between customer $i$ and item $j$ with probability $p$; and (c) for each edge $(i,j)$ in the preference graph, assign rating $+1$ with probability $\alpha_k$ and $-1$ with prob. $1-\alpha_k$ where customer $i$ belongs to segment $k$.

{\bf LC benchmark.} 
Given the customer-item preference graph $\Pcal$ with the corresponding ratings $\bmx_1, \bmx_2,\dotsb, \bmx_m$, the LC method estimates the model parameters by solving the following MLE problem:
\begin{equation*}
	\max_{\substack{q_1, q_2,\dotsc, q_K \\ \alpha_1,\dotsc, \alpha_K}} \; \sum_{i=1}^m \log\left( \sum_{k=1}^K q_k \prod_{j \in N(i)} \alpha_{k}^{\ind[x_{ij} = +1]} {(1-\alpha_{k})}^{\ind[x_{ij} = -1]}\right) \text{ s.t. } \sum_k q_k = 1, q_k \geq 0, 0 \leq \alpha_{k} \leq 1~\forall~k
\end{equation*}
The above optimization problem is non-concave and in general hard to solve to optimality. To address this computational challenge, the LC method adopts the popular Expectation Maximization (EM) heuristic by treating the segment membership of each customer $i$ as a latent variable $z_i$ (refer to Appendix~\ref{app:lc-em} for more details).
After the model parameters are obtained, customers are assigned to the segment for which the posterior probability of membership is the largest. Note that the LC method estimates a total of $2\cdot K$ parameters. 

{\bf Model-based projection algorithm.} We implement the model-based projection technique for segmenting customers as outlined in Algorithm~\ref{alg:gen_seg}.
In particular, the pooled model $f_{\pool}$ is just a single parameter, which we refer to as $\alphapool$:
\[ \alphapool = \frac{\sum_{(i,j) \in \Pcal} \ind[x_{ij} = +1]}{\abs{\set{(i,j) : (i,j) \in \Pcal}}} \]
We cluster the projection scores using $k$-means to obtain the segments, and call this method~{\sf proj}.
\begin{table}
\TABLE
{Percentage accuracy in recovering true segments for different parameter settings.\label{tab:simul_acc}}
{\tabcolsep=12.6pt
\begin{tabular}[t]{*{5}{c}} \toprule
$K$ & $1-p$ & LC & {\sf proj} & {\sc \% improvement} \\
\midrule
\multirow{5}{*}{5} & 0.0 & 99.0 & 98.7 & -0.3 \\
 & 0.2 & 98.1 & 97.5 & -0.6 \\
 & 0.4 & 96.2 & 95.1 & -1.1 \\
 & 0.6 & 91.8 & 89.2 & -2.8 \\
 & 0.8 & 75.7 & 75.9 & $0.3^{*}$ \\
 \cmidrule(lr){2-5}
 Avg. running time (secs) & & 0.45 & 0.04 & \\
\midrule
\multirow{5}{*}{7} & 0.0 & 92.9 & 92.1 & -0.9 \\
 & 0.2 & 89.3 & 88.3 & -1.1 \\
 & 0.4 & 82.5& 83.1 & $0.7^{*}$\\
 & 0.6 & 71.9 & 74.6 & 3.8\\
 & 0.8 & 54.2& 61.4 & 13.3\\
 \cmidrule(lr){2-5}
Avg. running time (secs) & & 0.52 & 0.05 & \\
\midrule
\multirow{5}{*}{9} & 0.0 & 72.5& 80.8& 11.4\\
 & 0.2 & 65.7& 75.6& 15.1\\
 & 0.4 & 58.3 & 70.4 & 20.7\\
 & 0.6 & 47.9& 61.5& 28.4\\
 & 0.8 & 39.1 & 49.1& 25.6\\
 \cmidrule(lr){2-5}
Avg. running time (secs) & & 0.53 & 0.08 &\\
\bottomrule 
\end{tabular}}
{The parameters are $K$---number of customer segments and $(1-p)$---sparsity of the preference graph. Each observation above is an average over $30$ experimental runs. All improvements are statistically significant according to a paired samples $t$-test at $1\%$ significance level, {\bf except} the ones marked with $^*$.}
\end{table}

{\bf Results and Discussion.} We measure the quality of the recovered clusters in terms of {\em accuracy}, defined as 
\[ {\rm Accuracy}^{\rm {\sf algo}} = 100 \times \left(\frac{1}{m} \sum_{i=1}^m \ind[z_i^{{\sf algo}} = z_i] \right) \]
where {\sf algo} $\in$ \{LC, {\sf proj}\}, $z_i$ is the true segment of customer $i$ and $z_i^{{\sf algo}}$ is the segment label assigned by method {\sf algo}. 
To account for permutations in the assigned segment labels, the true segments are ordered such that $\alpha_1 < \alpha_2 < \dotsb < \alpha_K$. Then, for the LC method, we assign the segment labels in order of the estimated alpha parameters $\hat{\alpha}_k$, so that $\hat{\alpha}_1 < \hat{\alpha}_2 < \dotsb < \hat{\alpha}_K$. For the {\sf proj} method, recall from Lemma~\ref{thm:lc_ind_conc} that the projection scores of customers in segment $k$ concentrate around $H(\alpha_k, \alphapool)/H(\alphapool)$. Since $H(\alpha_k, \alphapool) = -\alpha_k \log \frac{\alphapool}{1-\alphapool} - \log (1-\alphapool)$, it follows that $H(\alpha_k, \alphapool)$ is either increasing or decreasing in $\alpha_k$ depending on whether $\alphapool < \frac{1}{2}$ or $> \frac{1}{2}$. Therefore, we assign the segment labels in the increasing (resp. decreasing) order of the customer projection scores when $\alphapool < \frac{1}{2}$ (resp. $\alphapool > \frac{1}{2}$).

Table~\ref{tab:simul_acc} reports the accuracy of the LC and {\sf proj} methods. Since there is no model misspecification, the LC method is statistically optimal and we see that it is able to recover the true customer segments accurately when the preference graph is dense, but its performance suffers as the sparsity, $1-p$, increases. This happens because the number of data points per customer scales with $p$ while the number of parameters estimated scales with $K$, so that the LC method encounters insufficient data to reliably estimate the model parameters as the sparsity increases.
The {\sf proj} method, on the other hand, has comparable performance when there is enough data relative to the number of parameters being estimated. Further, as the level of sparsity increases, so that lesser data 
per customer is available, it outperforms the LC benchmark, with improvements upto $28\%$. This is because the {\sf proj} method estimates only a single parameter and therefore, can make more efficient use of the available customer labels to determine the true segments.

Finally, as reported in Table~\ref{tab:simul_acc}, the fact that we estimate only a single model as opposed to $K$ models results in upto $11\times$ gains in the running time compared to the LC benchmark, which is also sensitive to the initialization of the model parameters. This speedup becomes more significant when we have millions of customers and items, such as in our case-study at eBay (see Section~\ref{sec:ebay}), where the LC method is too computationally expensive and becomes infeasible in practice.

\newcommand{\jnew}{j_{\rm new}}
\newcommand{\gitrain}{N^{\rm train}(i)}
\newcommand{\gitest}{N^{\rm test}(i)}
\newcommand{\amle}{\alphapool}
\newcommand{\hzi}{\hat{z}_i}
\newcommand{\akmle}{\alphak^{\rm proj}}
\newcommand{\mathlca}{{\rm LC}}
\newcommand{\mathpop}{{\rm {\sf pop}}}
\newcommand{\mathproj}{{\rm {\sf proj}}}
\newcommand{\rpop}{\hat{r}_{i\jnew}^\mathpop}
\newcommand{\rlca}{\hat{r}_{i\jnew}^\mathlca}
\newcommand{\rpen}{\hat{r}_{i\jnew}^\mathproj}

\section{Case study 1: Cold start recommendations in MovieLens dataset}
\label{sec:movielens}
In this section, we focus on solving the classical cold-start problem in recommendation systems---recommending {\em new} movies to users\footnote{We refer to customers as ``users'' in the remainder of the paper}, in the context of the popular {\tt MovieLens}~\citep{herlocker1999algorithmic} dataset. This problem is challenging because by definition, {\em new} movies do not have any existing ratings from users, and still we need to determine which users the movie should be recommended to. We show that segmenting the user population using our approach and providing customized recommendations to each segment can result in upto $48\%$, $55\%$ and $84\%$ improvements in the recommendation quality for drama, comedy and action movies respectively; when compared to treating the user population as having homogeneous preferences and recommending the same movies to all users. In addition, it also outperforms the standard LC benchmark by $8\%$, $13\%$ and $10\%$ respectively while achieving a $20\times$ speedup in the computation time.

The cold-start problem~\citep{schein2002methods} has been studied extensively in the recommendation systems literature with solutions utilizing user-level and item-level attributes~\citep{park2009pairwise,zhang2014addressing} as well as social connections such as Facebook friends/likes or Twitter followers~\citep{lin2013addressing,sedhain2014social}. However the {\tt MovieLens} dataset only consists of user ratings on a 1-5 scale, and the genre of the movies. Consequently, we consider the following setup for the cold-start problem:

{\bf Setup}. Our goal is to recommend to users new movies that they will like, so we pose this recommendation task as the following prediction problem: given a movie, what is the probability that the user {\em likes} the movie? We say that a user {\em likes} a movie if the rating for the movie is greater than or equal to the average rating of the user across all rated movies, and {\em dislikes} the movie otherwise. Since the prediction task is only concerned with a binary (like/dislike) signal, we transform the raw user ratings to a binary like (+1) and dislike (-1) scale.
Given this, the goal becomes to predict the probability that a user gives $+1$ rating to a movie. We solve the prediction problem separately for each genre, since user preferences can vary across different genres. However, to keep the notation succinct, we do not explicitly denote the dependence on the genre in the following discussion.


To determine the benefits of segmentation for solving this prediction problem, we contrast two approaches---(1) {\em population model} : the user population is homogeneous so that all users have the same probability $\alpha$ for liking any movie; and (2) {\em segmentation model} : the population is composed of $K$ segments, such that users in segment $k$ have probability $\alphab_k$ of liking any movie. 
Given ratings from a population of users for a collection of movies, called the {\em training} set, we first estimate the model parameters in both approaches. Then, based on the estimated parameters, we predict the ratings given by each user for movies in a hold-out (or {\em test}) set. Let $U$ denote the set of all users, and $\gitrain$ and $\gitest$ denote respectively the set of movies in the training and test set rated by user $i$. For the {\em population model} approach, the maximum likelihood estimate (MLE) for parameter $\alpha$ is obtained by pooling all the ratings:
\[ \alphapool = \frac{\sum_{i \in U} \sum_{j \in \gitrain} \ind[r_{ij} = +1]}{\sum_{i \in U} \abs{\gitrain}} \]
where $r_{ij}$ is the rating given by user $i$ for movie $j$. For the {\em segmentation model} approach, we compare two algorithms: latent class (LC) and our model-based projection technique {\sf proj}, described earlier in Section~\ref{sec:simul_exp}. 
Let $\alphak^{\rm LC}$, $k=1,2,\ldots, K$, denote the segment parameters estimated by the LC method. The {\sf proj} method computes the user projection scores based on the estimated pooled model $\alphapool$. Once we obtain the different user segments, the {\sf proj} method computes each segment parameter as:
\[ \akmle = \frac{\sum_{i \in U} \ind[\hzi^\mathproj = k] \cdot \left( \sum_{j \in \gitrain} \ind[r_{ij} = +1]\right)}{\sum_{i \in U} \ind[\hzi^\mathproj = k] \cdot \abs{\gitrain}} \]
where $\hzi^\mathproj \in [K]$ represents the assigned segment label for user $i$.

Given the above, the prediction for user $i$ and new movie $\jnew$ is carried out as follows:
\begin{enumerate}
\item For the {\em population model}, the predicted rating $\rpop = +1$ if $\amle \geq 0.5$ otherwise $\rpop = -1$.
\item For the {\em segmentation model}, first consider the LC method. Let $\gamma_{ik}$ denote the posterior probability of membership in segment $k$ for user $i$. Then, $\rlca = +1$ if $\sum_{k=1}^K \gamma_{ik} \alphak^{\rm LC} \geq 0.5$ else $\rlca = -1$. For the {\sf proj} method, we have $\rpen = +1$ if $\alpha_{\hzi^\mathproj}^{\rm proj} \geq 0.5$ else $\rpen = -1$.
\end{enumerate}

\newcommand{\method}{{\sf method}}
There are many metrics to evaluate recommendation quality~\citep{shani2011evaluating}. Since we are dealing with binary ratings, a natural metric is {\em accuracy}, i.e. the fraction of ratings that are predicted correctly. More precisely, let $U^{\rm test}$ denote the set of all users in the test set, note that $U^{\rm test} \subseteq U$. Then for a user $i \in U^{\rm test}$, we define the following: 
\begin{equation*}
{\rm Accuracy}^{\rm \method}_i = \frac{1}{|\gitest|}\sum_{\jnew \in \gitest} \ind[r_{i\jnew} = \hat{r}_{i\jnew}^{\rm \method}] 
\end{equation*}
where ${\rm \method} \in \{{\sf pop}, {\rm LC}, {\sf proj}\}$.
The aggregate accuracy is then computed as:
\[ {\rm Accuracy}^{\rm \method} = 100 \times \left(\frac{1}{\abs{U^{\rm test}}} \sum_{i \in U^{\rm test}} {\rm Accuracy}^{\rm \method}_i\right) \]
In the same manner, we can also compute the aggregate accuracy for a given segment $k$ of users (identified by the {\sf proj} method):
\[ {\rm Accuracy}_k^{\rm \method} = 100 \times \left(\frac{1}{\abs{\set{i \in U^{\rm test} : \hzi^\mathproj = k}}} \sum_{i \in U_{\rm test} : \hzi^\mathproj = k} {\rm Accuracy}^{\rm \method}_i\right) \]

\begin{figure}
\FIGURE
{\begin{minipage}[c]{0.33\linewidth}
\includegraphics[scale=0.2]{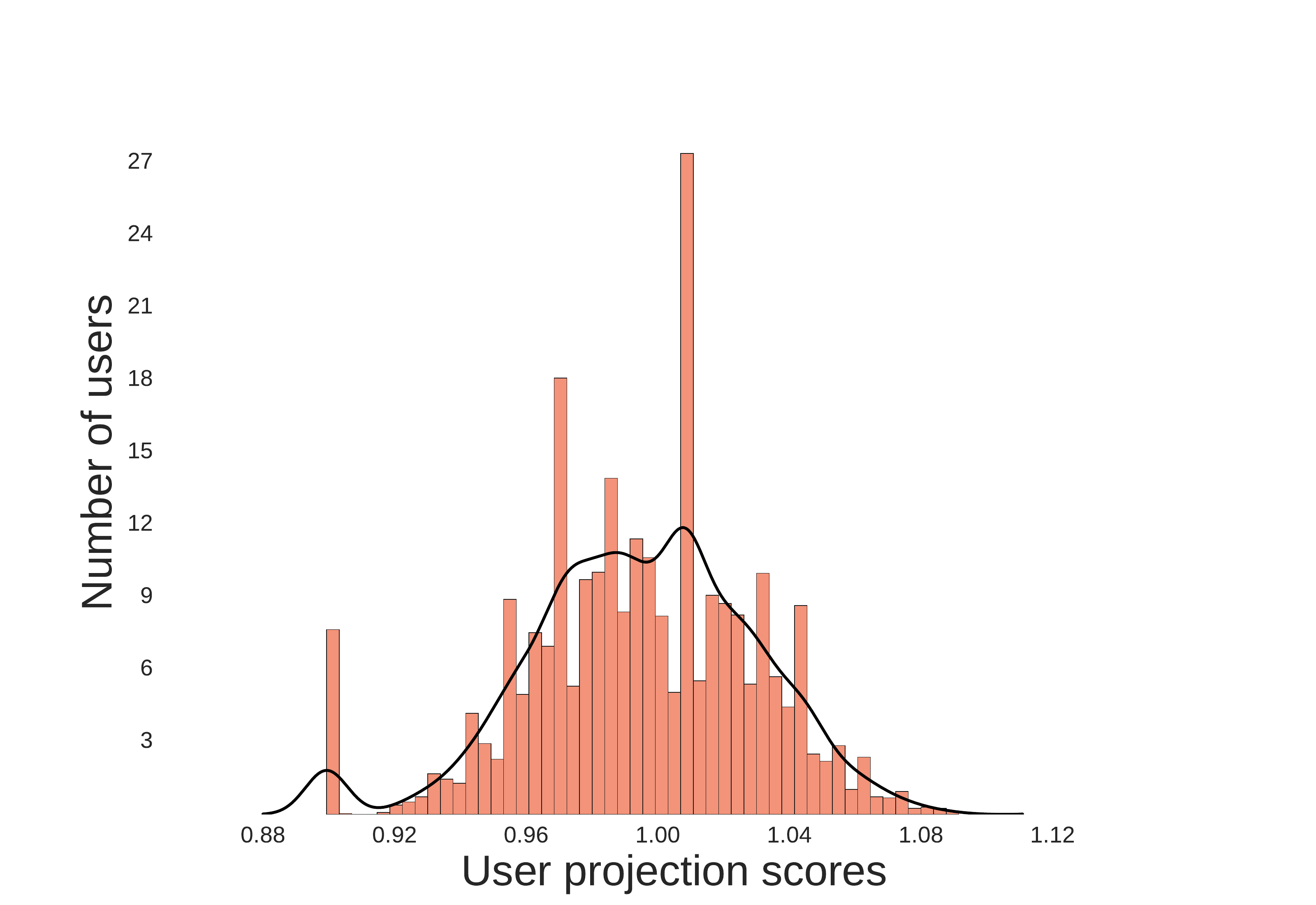}
\end{minipage}
\begin{minipage}[c]{0.33\linewidth}
\includegraphics[scale=0.2]{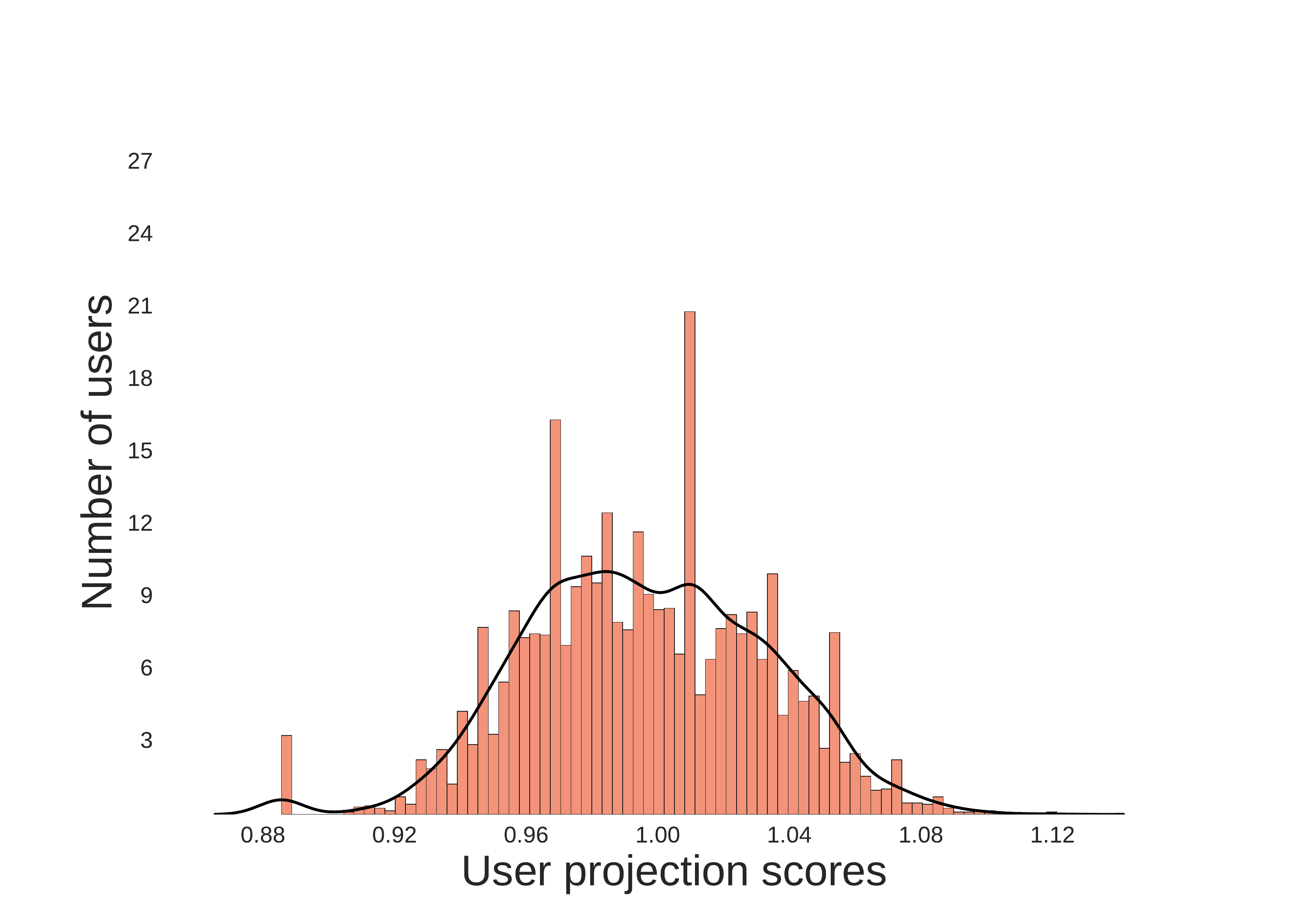}
\end{minipage}
\begin{minipage}[c]{0.33\linewidth}
\includegraphics[scale=0.2]{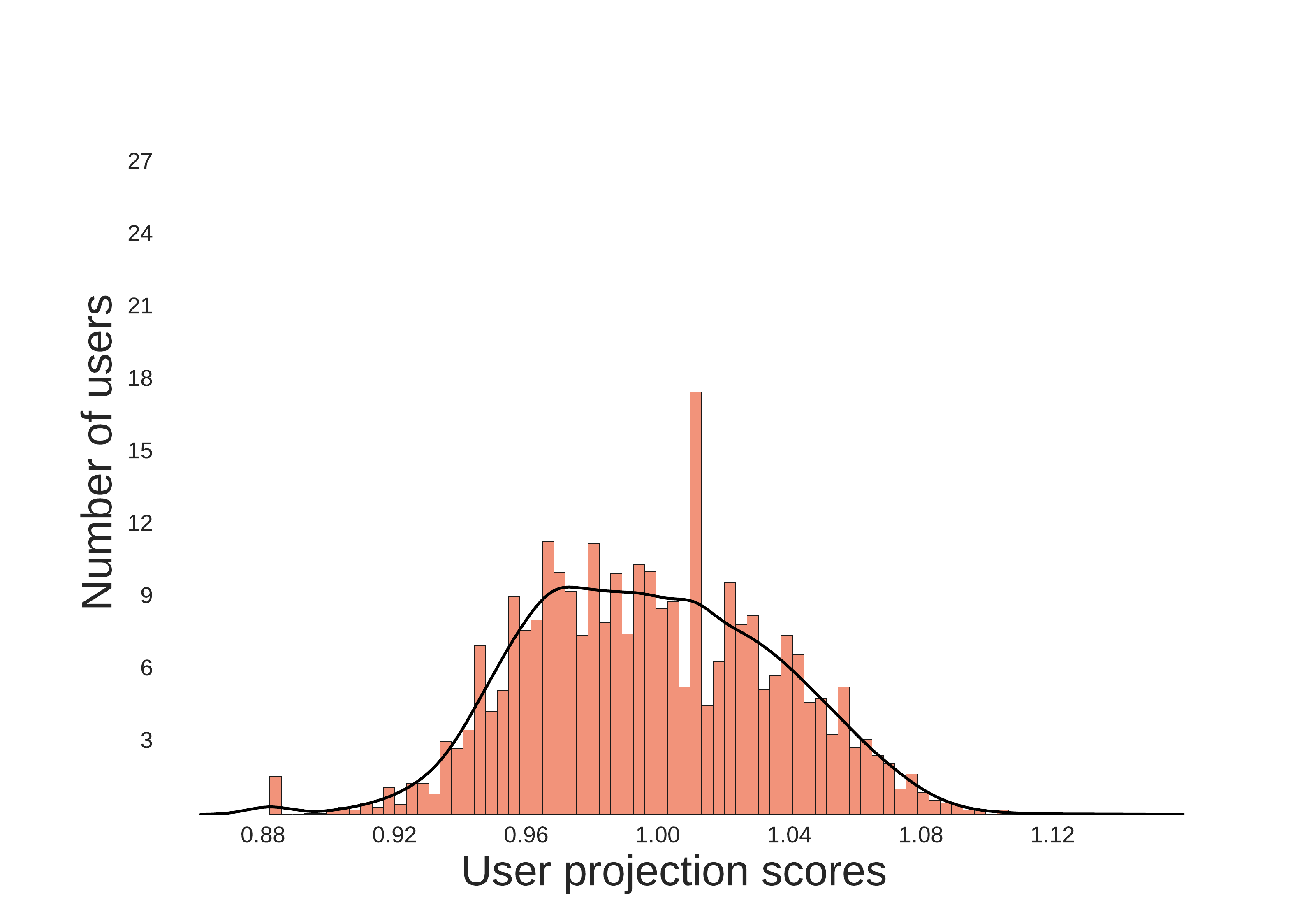}
\end{minipage}
}
{Density of user projection scores for the different genres -- {\bf Left}: Action, {\bf Center}: Comedy and, {\bf Right}: Drama.\label{fig:user-proj}}
{}
\end{figure}

\begin{table}
\TABLE
{Training/test data statistics and aggregate rating prediction accuracy for the different genres.\label{tab:ml-stats}}
{\tabcolsep=3.2pt
\begin{tabular}[t]{*{11}{c}} \toprule
Genre & \multicolumn{2}{c}{Train Data} & \multicolumn{2}{c}{Test Data} & \multicolumn{3}{c}{Accuracy} & \multicolumn{3}{c}{\% increase} \\
\cmidrule(lr){2-3} \cmidrule(lr){4-5} \cmidrule(lr){6-8} \cmidrule(lr){9-11}
 & \mcOne{Users} & \mcOne{Movies} & \mcOne{Movies} & \mcOne{Ratings} & \mcOne{${\rm {\sf pop}}$} & \mcOne{${\rm LC}$} & \mcOne{${\rm {\sf proj}}$} & \mcOne{${\rm LC}$ over $\mathpop$} & \mcOne{${\rm {\sf proj}}$ over $\mathpop$} & \mcOne{$\mathproj$ over LC}\\
\midrule
Action (K=2) & 6012 & 453 & 42 & 403 & 30.7 & 51.2 & 56.4 & 66.8 & 83.7 & 10.1\\
Comedy (K=4)& 6031 & 945 & 218 & 2456 & 37.7 & 51.8 & 58.4 & 37.4 & 54.9 & 12.7 \\
Drama (K=4) & 6037 & 1068 & 425 & 4627 & 38.6 & 53.0 & 57.2 &  37.3& 48.2 & 7.9\\
\bottomrule 
\end{tabular}}
{The number in brackets represents the number of segments determined for each genre.}
%
\end{table}

{\bf Results and Discussion}. The {\tt MovieLens} dataset consists of 1 million movie ratings from 6,040 users for 3,952 movies. For our analysis, we choose the three most popular genres in the dataset---drama, comedy and action, and consider movies (in each genre) that have been rated by atleast 30 users as part of the training set, and all other movies as part of the test set. Statistics for the training and test datasets are given in Table~\ref{tab:ml-stats}. As discussed earlier in Section~\ref{sec:lreg_graph}, our approach can provide data-driven guidance in choosing the number of segments based on the number of peaks in the estimated density of the user projections scores. Figure~\ref{fig:user-proj} shows the kernel density estimate of the projection scores for each genre; we try
values of $K \in \set{2,3,4,5}$ and choose the best using cross-validation. For the LC method, we choose the best of 10 random initializations. After segmenting the users, we predicted their ratings for new movies as outlined above and compute the accuracy metrics for the different approaches. 

Table~\ref{tab:ml-stats} reports the aggregate accuracy for each of the genres. The benefits of segmentation can be seen across all the genres, with improvements upto $84\%$ (for the action genre) in the prediction accuracies. The population model treats the preferences of all users as being the same and performs poorly since it ends up recommending the same set of movies to all the users. 
The segmentation model, on the other hand, makes different recommendations to the different user segments, and consequently performs significantly better. 
Further, using our segmentation algorithm performs better than the LC method (upto $13\%$ for the comedy genre) showing that a ``hard'' separation of the users into distinct segments is better than a ``soft'' clustering, where users have membership in multiple segments, for the cold-start recommendation problem. In addition, the {\sf proj} method is upto 20x faster than the LC method when the population is grouped into $K=4$ segments, again highlighting the fact that our algorithm is fast and can scale to large dimensional data.

\begin{table}
\TABLE
{Comparison of rating prediction accuracy of population model and our model-based projection technique by individual user segments.\label{tab:pred-metrics}}
{\tabcolsep=9.6pt
\begin{tabular}[t]{*{7}{c}} \toprule
Genre & $\alphapool$ & User Segments & $\alpha_k^{\rm {\sf proj}}$ & ${\rm Accuracy}_k^{\rm {\sf pop}}$ & ${\rm Accuracy}_k^{\rm {\sf proj}}$ & $\%$ increase \\ 
\midrule
\multirow{2}{*}{Action} & \multirow{2}{*}{0.537} & Segment 1 (2900)	& 0.658 & 35.6	& 35.6	&- \\ 
& & Segment 2 (3112)	& 0.441 & 26.9	& 73.1	& 171.7 \\
\midrule		
\multirow{4}{*}{Comedy} & \multirow{4}{*}{0.543} & Segment 1 (941)& 0.749 & 45.2 & 45.2&-  \\	
& & Segment 2 (2062)		& 0.631 & 45.9  &45.9 & - \\
& & Segment 3 (1869)	& 0.495 & 33.2	&66.8  &101.3 \\
& & Segment 4 (1159)	& 0.360 & 26.4	&73.6  &178.7 \\
\midrule
\multirow{4}{*}{Drama} & \multirow{4}{*}{0.545} & Segment 1 (1164)	& 0.736 & 50.1 &	 50.1 & - \\ 
& & Segment 2 (1975)	& 0.620 & 43.5 & 43.5 & -			\\ 
& & Segment 3 (1769)	& 0.485 & 36.1	& 63.9 & 77.0 		\\
& & Segment 4 (1129)	& 0.342 & 22.5	 & 77.5 & 244.4	\\

\bottomrule 
\end{tabular}}
{The numbers in the bracket represent the size of each user segment. $\%$ increase denotes the percentage improvement in accuracy of our segmentation approach over the population model}
\end{table}

To understand where the accuracy improvements come from, Table~\ref{tab:pred-metrics} displays the accuracy of the {\sf pop} and {\sf proj} methods, broken down by individual user segments computed by the {\sf proj} method. Also shown are the estimated pooled model $\alphapool$ and segment parameters $\alpha_k^{\rm {\sf proj}}$. Now observe that for segments 1 \& 4 in the drama and comedy genres, the estimated parameters $\alpha_k^{\rm {\sf proj}}$ are furthest from the pooled parameter $\alphapool$. In other words, these segments contain users whose preferences are very different from that of the population, i.e. {\em esoteric} preferences, which are not captured well by the pooled model $\alphapool$. Using the segment parameters $\alpha_k^{\mathproj}$ for the rating predictions results in significant improvements in the accuracy for segment 4 users---upto 1.8x and 2.5x increase for the comedy and drama genres respectively. Note that we do not observe any improvement for segment 1 users, this is because of our experimental setup which involves a coarse-grained rating prediction based on a threshold of 0.5 (see the setup above). The users in the intermediate segments 2 \& 3, on the other hand, have preferences that are very similar to those of the population, i.e {\em mainstream} preferences. However, we are still able to distinguish between users in these segments resulting in improved rating prediction accuracy for segment 3 users. The improvements are lower than for the esoteric segments, since the pooled model is already able to capture the preferences of the mainstream users. The story is similar for the action genre where segment 1 (resp. 2) behaves as the mainstream (resp. esoteric) segment.

To summarize, our model-based projection technique for segmenting users can be used to generate high-quality personalized new movie recommendations. As pointed out earlier, two of the benefits of our algorithm are its {\em flexibility}, so that it can be applied in different application contexts without much customization, and {\em scalability}, as we showed above, our approach is an order of magnitude faster than a standard latent class approach. We further illustrate these benefits in the context of recommending similar items at eBay, a large e-commerce retailer.

\section{Case study 2: Personalized Recommendations on eBay}
\label{sec:ebay}

\begin{figure}
\FIGURE
{\includegraphics[scale=0.4]{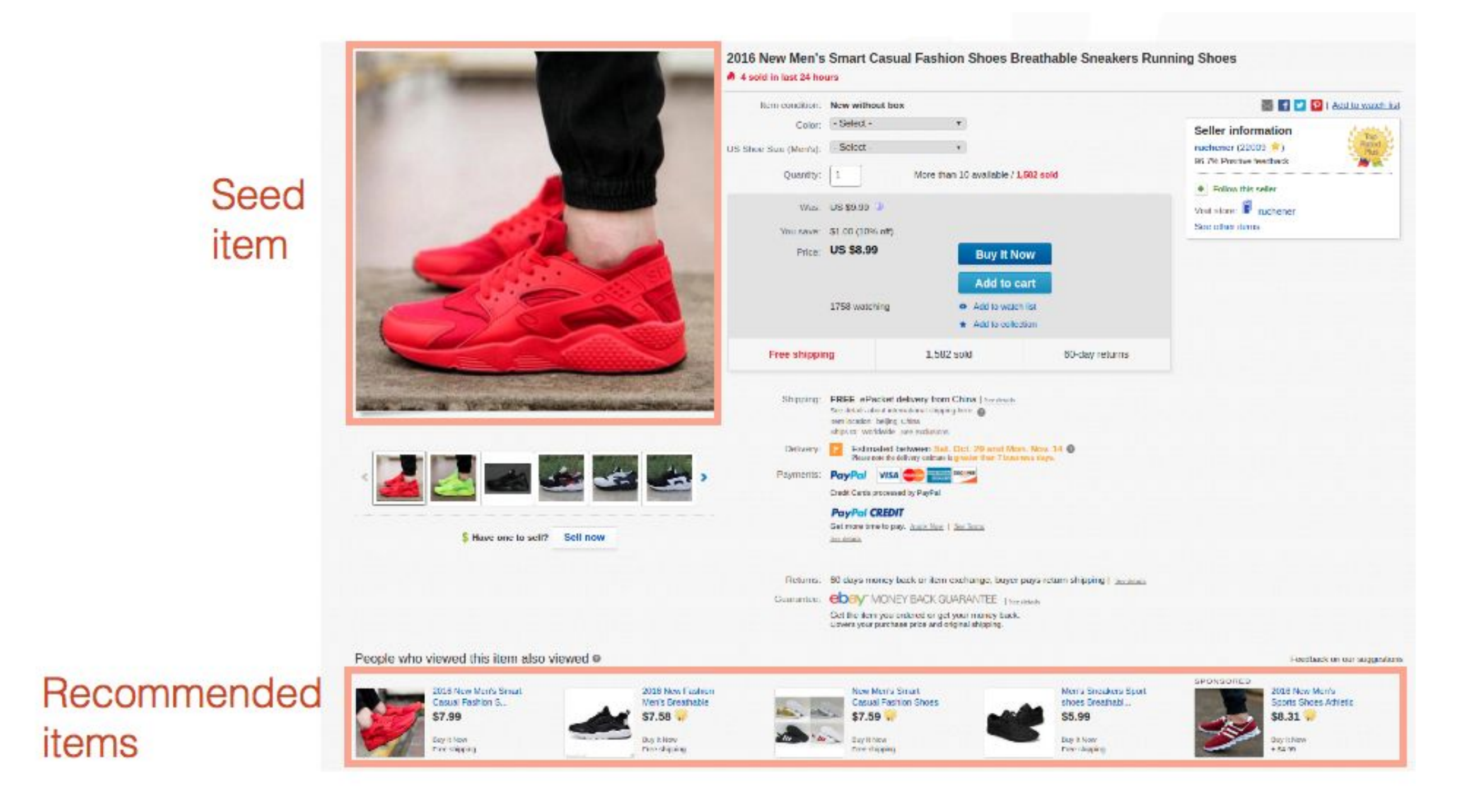}
}
{Example of similar item recommendation on eBay.\label{fig:ebay-ex}}
{The seed item is a pair of sneakers that the user is currently viewing, and the recommended items at the bottom are other pairs of shoes similar to the seed item}
\end{figure}

In this section, we describe the application of our segmentation methodology for personalizing similar item recommendations on eBay. Specifically, the goal is to recommend items ``similar'' to a given {\em seed} item, which is an item that a user is currently viewing, see Figure~\ref{fig:ebay-ex} for an example. The recommended items are shown below the seed item, above the fold.\footnote{Above the fold refers to the portion of the webpage that is visible without scrolling} This problem is complex since the eBay marketplace offers diverse listings ranging from Fitbit tracker/iPhone (products with well-defined attributes) to obscure antiques and collectibles which may be highly unstructured. This is compounded by the fact that the listings can have multiple conditions---new, used, refurbished, etc.---and selling formats---fixed price vs auction. In addition, most users might interact with only a small fraction of the catalog---in our sample dataset below, a user on average interacted with 5 out of 2M items---which makes it hard to determine their preferences and limits the application of traditional collaborative filtering algorithms.~\cite{Brovman:2016} designed a scalable recommendation system at eBay with the aim of addressing these challenges. While their approach resulted in positive lift in critical operational metrics, it does not take into account heterogeneous user preferences---for a given seed item, {\em every} user is recommended the {\em same} set of items. We show that segmenting the user population based on our technique can result in upto $6\%$ improvement in the recommendation quality. This improvement is non-trivial because eBay also tried several natural segmentations such as similarity in demographics (age, gender, income) or aggregate purchase behavior like the number of transactions or amount spent in the last year, etc. but all of them resulted in $< 1\%$ improvement.

{\bf Segmenting users based on their click and purchase activity}.  
The existing approach of Brovman et al. does not consider heterogeneous preferences, which are manifested in the browsing and purchase behavior of users on eBay---for instance, some users might always purchase used (or refurbished) items, whereas others like to place bids for items on auction to obtain the best~prices.

We applied our model-based projection technique to segment the user population and provide personalized similar item recommendations. We first create the preference graph as follows: there is a node for each user and each recommended item. For a given user $i$ and some recommended item $j$ that was shown to the user, we encode the observation as $x_{ij} \in \set{\text{non-click}, \text{click}, \text{purch}}$ based on whether the user did not click on $j$, clicked on $j$ but did not purchase $j$, or clicked and then purchased item $j$. We only consider user activity on these items when they are part of the recommendations for any seed item, and ignore other interactions on the eBay site. In the original paper, the authors used a combination of {\em comparison} features like price, condition, selling format, etc. which compare attributes of the seed and recommended items, as well as {\em item quality} features like seller feedback which capture the intrinsic quality of a recommended item. However, because of the large diversity and quantity of items on eBay, many of the recommended items did not have well-defined values for some of these features.
%
Therefore, we chose to estimate the empirical distribution of the raw user observations, as follows: let $\djplus, \djzero$ and $\djminus$ denote respectively the number of users that clicked and purchased, clicked but did not purchase and, did not click some recommended item $j$. Then we estimate the pooled model as $f_{\pool}(\bmx) = \prod_{j=1}^n f_{\pool,j}(x_j)$ where each $f_{\pool,j} : \set{\text{non-click}, \text{click}, \text{purch}} \to [0,1]$ is such that 
\[ f_{\pool,j}(\text{label}) = \frac{d_j^{\text{label}}}{\djplus + \djminus + \djzero}  \]
where $\text{label} \in \set{\text{non-click}, \text{click}, \text{purch}}$. Having estimated the pooled model, we segment the user population by $k$-means clustering of the scores computed by Algorithm~\ref{alg:lreg_seg}.

\begin{figure}
\FIGURE
{\begin{minipage}[c]{0.50\linewidth}
\includegraphics[scale=0.34]{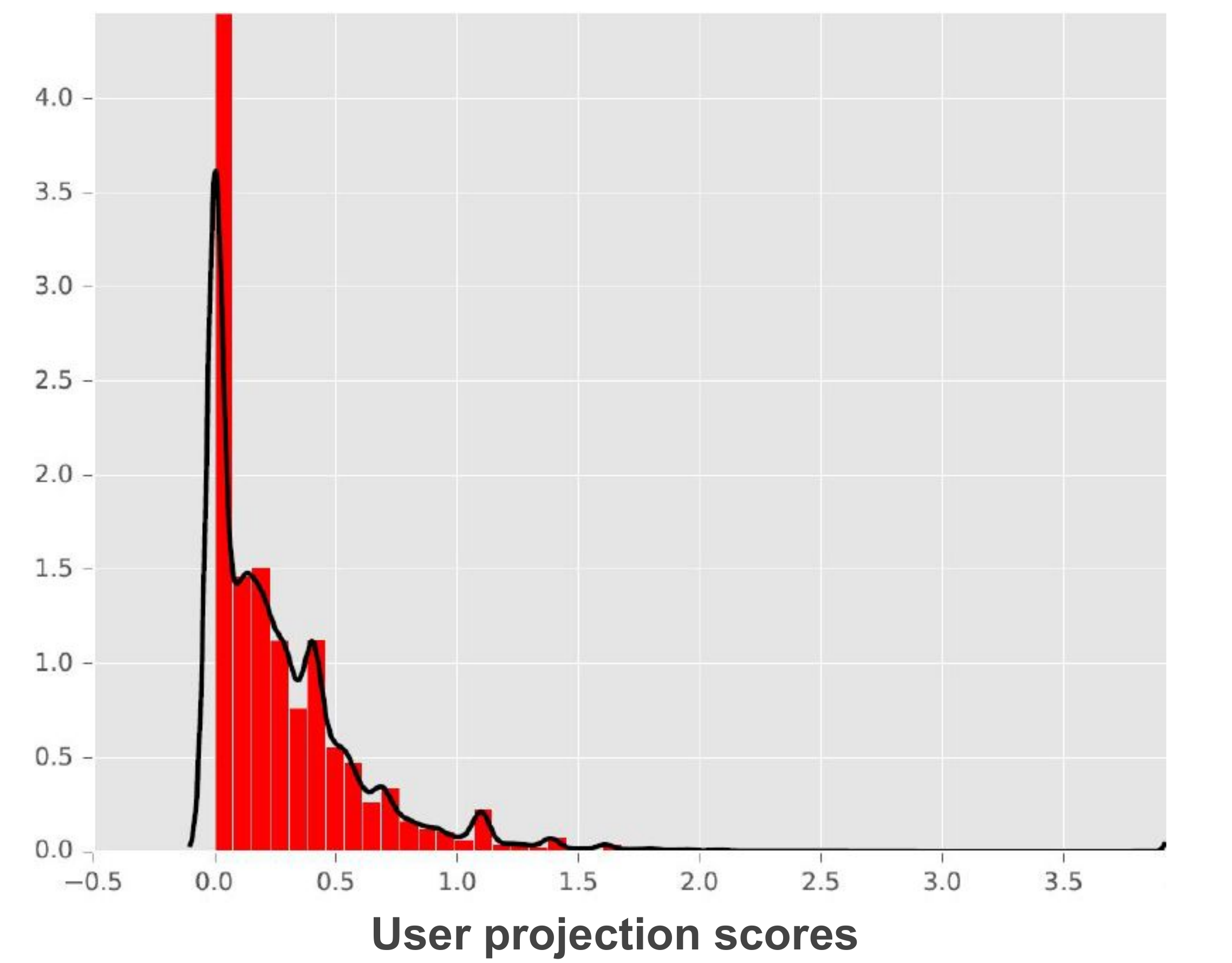}
\end{minipage}
\begin{minipage}[c]{0.48\linewidth}
\includegraphics[scale=0.35]{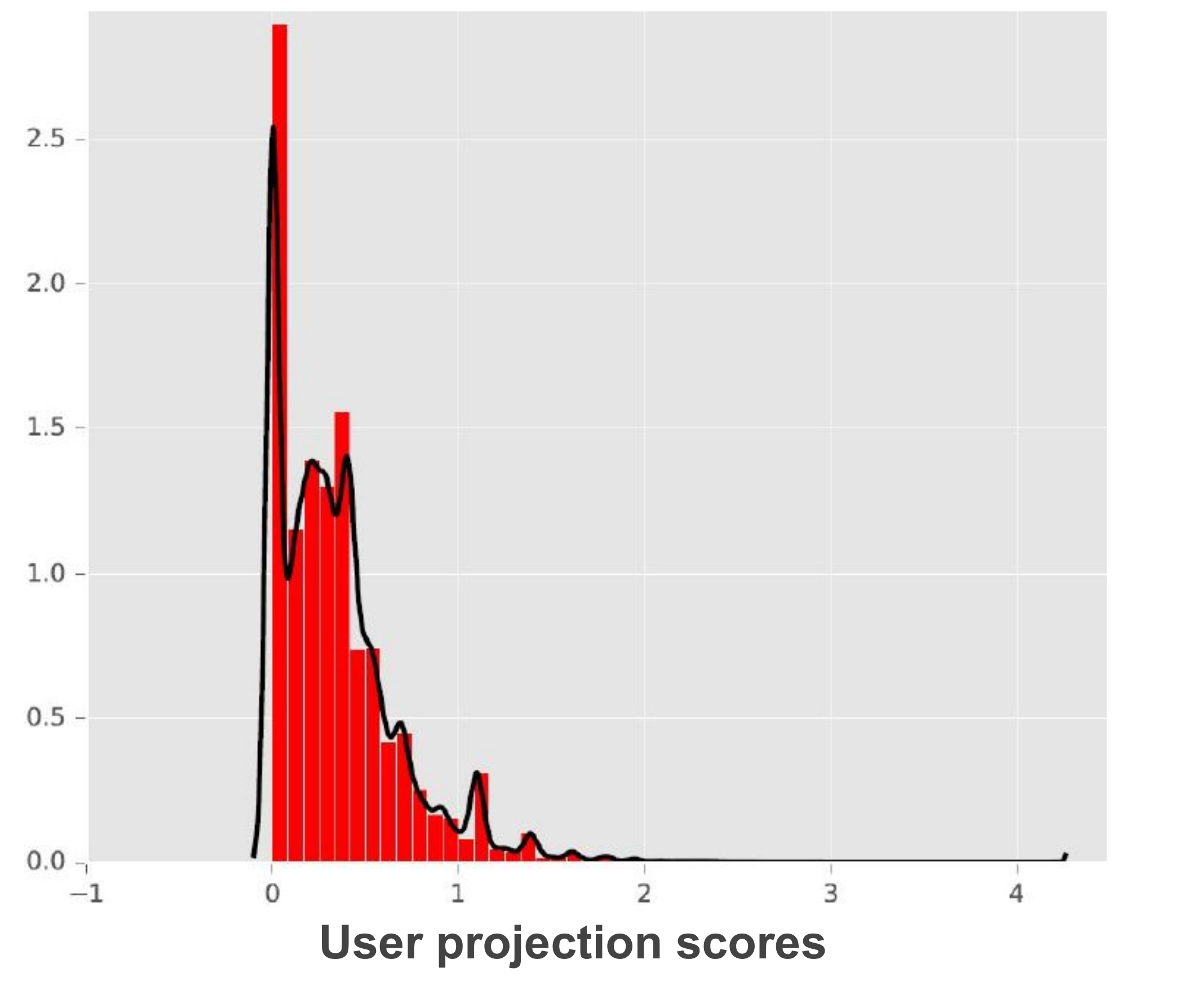}
\end{minipage}}
{Density of user projection scores for two product categories, {\bf Left}: Clothing, Shoes and Accessories and {\bf Right}: Home \& Garden\label{fig:ebay-pen}}
{}
\end{figure}

{\bf Results and Discussion}. We consider a dataset of 1M users along with their click and purchase activity on roughly 2M recommended items over a 2 week period. Since eBay has many different product categories, Brovman et al. performed product category segmentation and we report the results for the two most popular categories---{\tt Clothing, Shoes and Accessories} and {\tt Home \& Garden}; we obtained qualitatively similar results in other categories. Figure~\ref{fig:ebay-pen} shows the distribution of the user projection scores computed by our algorithm. Based on the number of peaks in these density curves, we chose $K=3$ user segments for both categories.

Brovman et al. transformed the problem of ranking candidate recommendations into the following (binary) prediction problem: for a fixed seed item, what is the probability that a given recommended item is purchased? The recommended items are then ranked based on the probability of being purchased (refer to the original paper for more details). We compare the area-under-the-curve (AUC)\footnote{AUC is a measure of classifier performance and is equal to the likelihood that the classifier will rank (based on the probability of positive class label) a randomly chosen positive example higher than a randomly chosen negative example} of this binary classifier on a hold-out sample for the original approach, which we term the {\em population model}, as well as the approach where a separate classifier is estimated for each user segment, the {\em segmentation model}. Table~\ref{tab:rec-imp} shows the percentage improvement in AUC of the segmentation model over the population model for individual user segments. 
We observe that the segmentation model performs better across the board, with improvements upto $6\%$ in both product categories.
Similar to the MovieLens case study earlier, the improvements are higher for users with esoteric preferences (segments 1 \& 3), since the population model is not able to capture the preferences of these users. Segment 2 consists of mainstream users, whose preferences are well captured by the population model, and therefore the improvement is lower.

\begin{table}
\TABLE
{Improvement in AUC of binary classifier for individual user segments.\label{tab:rec-imp}}
{\tabcolsep=9.6pt
\begin{tabular}[t]{*{3}{c}} \toprule
Product Category & User Segments & $\%$ increase in AUC\\ 
\midrule
\multirow{3}{*}{Clothing, Shoes and Accessories} & Segment 1 (26k)& 6.0 \\
& Segment 2 (18k)		& 0.8 \\
& Segment 3 (3k)	& 6.0 \\
\midrule
\multirow{3}{*}{Home \& Garden} & Segment 1 (22k)	& 5.7  \\
& Segment 2 (17k)	& 0.8 \\
& Segment 3 (4k)	& 2.2	\\
\bottomrule 
\end{tabular}}
{The numbers in the bracket represent the size (in thousands) of each user segment.}
\end{table}


The $6\%$ improvement obtained using our segmentation technique is non-trivial considering the fact that eBay also tried several natural segmentations such as similarity in demographics (age, gender, income) or aggregate purchase behavior like the number of transactions or amount spent in the last year, etc. but all of them resulted in $< 1\%$ improvement in the AUC. Such demographic-based segmentation implicitly assumes that similarity in demographics or aggregate purchase behavior implies similarity in preferences, which might not be the case in practice. Instead, focusing on actual user activity such as clicks and purchases of individual items can help to directly capture their preferences. However, a major challenge in using such data is that it is extremely sparse, for instance, in the dataset above, users had  only 5-6 observations on average and consequently, most of the users do not have any overlap in the observations that they generate. This makes it hard to determine whether two users have similar preferences. Further, existing techniques like the LC method are prohibitively slow for such a large dataset. Our method ran in about 20 minutes on a single core without any optimizations, and because the above approach treated different items independently, it can be easily ported to large-scale distributed data processing frameworks like Spark\footnote{\url{spark.apache.org}}. This shows that our segmentation technique can scale to large datasets and work directly with fine-grained user observations (such as clicks and purchases on individual items) to improve the quality of personalized item recommendations.

\section{Conclusions}
\label{sec:conclusions}

This paper presents a novel method to segment customers based on their preferences. Our method is designed to incorporate observations from diverse data sources such as purchases, ratings, clicks, etc. as well as handle missing observations. We propose a ``model-based'' projection technique that makes use of probabilistic models to transform the customer observations into a consistent and comparable scale, and then deals with the missing data issue by carefully projecting the transformed data into a lower dimensional space. The projections are then clustered to determine the customer segments. Our technique builds upon existing ideas in the machine learning literature for clustering observations from a mixture model (such as Gaussian mixtures) and extends them to handle categorical data, as well as missing entries in the observations. Our method can also leverage the existing literature in marketing that has proposed rich models to capture detailed and fine-grained customer preference structures.
A key feature of our segmentation algorithm is that it is analytically tractable, and we derive precise necessary and sufficient conditions in order to guarantee asymptotic recovery of the true customer segments. Experiments on synthetic data show an improvement in the accuracy of recovering the true segments, over the standard EM-based benchmark, in conjunction with an order of magnitude speedup. Further, using two case studies, including a real-world implementation on eBay data, we show that our segmentation approach can be used to generate high-quality personalized recommendations.

There are a few natural directions and opportunities for future work. We focused on categorical data in this paper---since most of the observations collected about customers from firms is categorical---but our methodology can also be applied directly to continuous data.
However, estimating continuous distributions (required for the pooled model) come with their own set of challenges and it will be interesting to explore how our algorithm performs in such scenarios.
From the analytical perspective, it will be interesting to determine other generative models (especially from the exponential family) for customer observations under which our algorithm can recover the true segments. For instance, we could consider mixtures of binary logit models where each item $j$ is represented as a vector $\bm{y}_j$ in some feature space $\mathcal{Y}$. Imposing suitable constraints on the space $\mathcal{Y}$ as well as defining appropriate missing data mechanisms in the customer observations will be important in this regard.
More broadly, the idea of separating customers based on their deviation from the population opinion can be applied in other domains such as text reviews, images or even audio/speech; to obtain interesting ``domain-specific'' notions of mainstream and esoteric opinions. Finally, it would be useful to test the effectiveness of our segmentation method in terms of standard marketing and/or economics-oriented performance measures such as customer value, profitability, loyalty, etc.

%
%
%


\ACKNOWLEDGMENT{
The results on the eBay dataset were achieved during an internship by Ashwin Venkataraman in collaboration with his mentors Yuri Brovman and Natraj Srinivasan, and the authors would like to thank them and eBay for their help and feedback.
}

{\small
\bibliographystyle{ormsv080} 
\bibliography{ref} 
}


\newpage
\setcounter{page}{1}
\renewcommand{\thepage}{A\arabic{page}}

\setcounter{equation}{0}
\renewcommand{\theequation}{A\arabic{equation}}

\setcounter{theorem}{0}
\renewcommand{\thetheorem}{A\arabic{theorem}}
\setcounter{lemma}{0}
\renewcommand{\thelemma}{A\arabic{lemma}}
\setcounter{figure}{0}
\renewcommand{\thefigure}{A\arabic{figure}}
\setcounter{section}{0}
\renewcommand{\thesection}{A\arabic{section}}
\renewcommand{\thefootnote}{\arabic{footnote}}
\setcounter{footnote}{0}

\begin{center}
{\large \bf A Model-based Projection Technique for Segmenting Customers} \\

{\large \sc Appendix} \\
\iftrue
{Srikanth Jagabathula} 

Leonard N. Stern School of Business, New York University, \\
44 West Fourth St., New York, NY 10012

sjagabat@stern.nyu.edu

{Lakshmi Subramanian \hspace{1cm} Ashwin Venkataraman}

Courant Institute of Mathematical Sciences, New York University, \\
251 Mercer Street, New York, NY 10012

\{lakshmi,ashwin\}@cs.nyu.edu
\fi

\end{center}

\newcommand{\cpd}{{\sc corr-pen (data-driven) }}
\newcommand{\cpo}{{\sc corr-pen }}
\newcommand{\bt}{\bm{T}}
\newcommand{\bc}{\bm{C}}
\newcommand{\bTheta}{\bm{\Theta}}
\newcommand{\Bin}{\mathrm{Bin}}
\newcommand{\fiplus}{\bm{F}_i^+}
\newcommand{\xl}{\frac{\bmX}{l}}
\newcommand{\xirem}{\bmX_{-i}^{+}}
\newcommand{\bmN}{\bm{N}}
\newcommand{\bmD}{\bm{D}}

\begin{APPENDICES}

\vspace{.2in}
We begin by proving some general statements about random variables, that will be used in the proofs later.
\Halmos \endproof

\begin{lemma}
\label{lem:conc-relation}
Let $\bm{X}_1, \bm{X}_2, \ldots, \bmX_r, \bm{Y}$ be a collection of non-negative random variables. Let $a_1, a_2, \ldots, a_r, b$ be positive constants. Then, given any $0 < \varepsilon < 1$ and any $1 \leq i \leq r$, we have that
\begin{align*}
(i) \Pr\expt{\abs{\frac{\bmX_i}{\bmY} - \frac{a_i}{b}} > \varepsilon\frac{a_i}{b}} &\leq \Pr\expt{\abs{\bmX_i - a_i} > \varepsilon'a_i} + \Pr\expt{\abs{\bmY - b} > \varepsilon'b} \\
(ii) \Pr\expt{\abs{\bmX_i\bmY - a_ib} > \varepsilon a_ib} &\leq \Pr\expt{\abs{\bmX_i - a_i} > \varepsilon'a_i} + \Pr\expt{\abs{\bmY - b} > \varepsilon'b} \\
(iii) \Pr\expt{\abs{\sum_{i=1}^r \bmX_i - \sum_{i=1}^r a_i} > \varepsilon \cdot \left(\sum_{i=1}^r a_i\right)} &\leq \sum_{i=1}^r \Pr\expt{\abs{\bmX_i - a_i} > \varepsilon a_i} \\
(iv) \Pr\expt{\sum_{i=1}^r \abs{\bmX_i - a_i} > \varepsilon \cdot \left(\sum_{i=1}^r a_i\right)} &\leq \sum_{i=1}^r \Pr\expt{\abs{\bmX_i - a_i} > \varepsilon  a_i}
\end{align*}
where $\varepsilon' = \varepsilon/3$.
\end{lemma}
\proof{Proof.}
{\bf Part (i)}. Let $\bmz_1 = \frac{\bmX_1}{\bmY}$. We prove the result by contradiction. Suppose $\bmz_1 > (1 + \varepsilon)\frac{a_1}{b}$. Then, $\bmX_1 > (1 + \varepsilon')a_1$ or $\bmY < (1 - \varepsilon')b$. If not, we have the following:
\begin{align*}
\bmX_1 \leq (1 + \varepsilon')a_1, \;\; \bmY \geq (1 - \varepsilon')b &\implies \frac{\bmX_1}{\bmY} \leq \frac{(1+\varepsilon')a_1}{(1-\varepsilon')b} \\
&\implies \bmz_1 \leq (1 + \varepsilon)\frac{a_1}{b}
\end{align*}
where the last implication follows from the fact that $\frac{1+ \varepsilon'}{1-\varepsilon'} \leq 1 + \varepsilon$ when $\varepsilon' = \varepsilon/3$ and $0 < \varepsilon < 1$. This is a contradiction. Therefore we have that,
\[ \Pr\expt{\bmz_1 > (1+\varepsilon)\frac{a_1}{b}} \leq \Pr\bexpt{\bmX_1 > (1 + \varepsilon')a_1 \bigcup \bmY < (1 - \varepsilon')b} \leq  \Pr\expt{\bmX_1 > (1 + \varepsilon')a_1} +  \Pr\expt{\bmY < (1 - \varepsilon')b} \]
where the last inequality follows from the union bound. An analogous argument establishes that
\[ \Pr\expt{\bmz_1 < (1 - \varepsilon)\frac{a_1}{b}} \leq \left(\Pr\expt{\bmX_1 < (1 - \varepsilon')a_1} + \Pr\expt{\bmY > (1 + \varepsilon')b}\right) \]
which uses the fact that $\frac{1- \varepsilon'}{1+\varepsilon'} \geq 1 - \varepsilon$ when $\varepsilon' = \varepsilon/3$ and $0 < \varepsilon < 1$.
Combining the above two arguments, we get
\[ \Pr\expt{\abs{\bmz_1 - \frac{a_1}{b}} > \varepsilon\frac{a_1}{b}} \leq \left(\Pr\expt{\abs{\bmX_1 - a_1} > \varepsilon'a_1} + \Pr\expt{\abs{\bmY -b} > \varepsilon'b}\right) \]

{\bf Part (ii)}. Let $\bmW_1 = \bmX_1 \bmY$. Suppose $\bmW_1 > (1 + \varepsilon)a_1 b$. Then, $\bmX > (1 + \varepsilon')a_1$ or $\bmY > (1 + \varepsilon')b$. If not, we have the following:
\begin{align*}
\bmX \leq (1 + \varepsilon')a_1, \;\; \bmY \leq (1 + \varepsilon')b &\implies \bmX_1\bmY \leq {(1+\varepsilon')}^2 a_1 b \\
&\implies \bmW_1 \leq (1 + \varepsilon) a_1 b
\end{align*}
where the last implication follows because $(1+\varepsilon')^2 \leq 1 + \varepsilon$ for $\varepsilon' = \varepsilon/3$ and $0 < \varepsilon < 1$. This is a contradiction. Therefore, 
\[ \Pr\expt{\bmW_1 > (1+\varepsilon)a_1b} \leq \Pr\bexpt{\bmX_1 > (1 + \varepsilon')a_1 \bigcup \bmY > (1 + \varepsilon')b} \leq  \Pr\expt{\bmX_1 > (1 + \varepsilon')a_1} +  \Pr\expt{\bmY > (1 + \varepsilon')b} \]

Combining with the symmetric case gives:
\[ \Pr\expt{\abs{\bmW_1 - a_1 b} > \varepsilon a_1b} \leq \left(\Pr\expt{\abs{\bmX_1 - a_1} > \varepsilon'a_1} + \Pr\expt{\abs{\bmY -b} > \varepsilon'b}\right) \]

{\bf Part (iii)}. Define $\bmz \defas \sum_{i=1}^r \bmX_i$ and $A \defas \sum_{i=1}^r a_i$. Suppose that $\bmz > (1+\varepsilon)A$. Then it follows that for some $1 \leq i \leq r$, $\bmX_i > (1+ \varepsilon)a_i$. If not, we have:
\begin{align*}
\bmX_1 \leq (1 + \varepsilon)a_1, \ldots \bmX_r \leq (1 + \varepsilon)a_r &\implies \sum_{i=1}^r \bmX_i \leq \sum_{i=1}^r (1+\varepsilon)a_i \\
&\implies \bmz \leq (1 + \varepsilon) A
\end{align*}
which is a contradiction. Combining with the symmetric case and applying the union bound, the claim follows.

{\bf Part (iv)}. Let $\bmz \defas \sum_{i=1}^r \abs{\bmX_i - a_i}$ and suppose $\bmz > \varepsilon \cdot \left(\sum_{i=1}^r a_i\right)$. Then it follows that for some $1 \leq i \leq r$, $\abs{\bmX_i - a_i} > \varepsilon a_i$. If not, we have:
\begin{align*}
\abs{\bmX_1 - a_1} \leq \varepsilon a_1, \ldots \abs{\bmX_r - a_r} \leq \varepsilon a_r &\implies \sum_{i=1}^r \abs{\bmX_i - a_i} \leq \sum_{i=1}^r \varepsilon a_i \\
&\implies \bmz \leq \varepsilon \cdot \left(\sum_{i=1}^r a_i\right)
\end{align*}
which is a contradiction. The claim then follows from the union bound.
\Halmos \endproof

\begin{lemma}
\label{lem:pen-conc}
Let $\bmX, \bmY$ be two non-negative random variables such that $0 \leq \bmX, \bmY \leq 1$ and $\bmY = 0 \implies \bmX = 0$. Suppose that $\E[\bmX], \E[\bmY] > 0$. Define the random variables $\bmz_1 = \bmX \cdot (-\log \bmY)$ and $\bmz_2 = \bmX \cdot (-\log \bmX)$, and the constants $A_1 = \E[\bmX] \cdot (-\log \E[\bmY])$ and $A_2 = \E[\bmX] \cdot (-\log \E[\bmX])$. Then, given any $0 < \varepsilon < 1$, we have:
\begin{align*}
\Pr\expt{\abs{\bmz_1 - A_1} > \varepsilon A_1} &\leq \Pr\expt{\abs{\bmX - \E[\bmX]} > \frac{\varepsilon}{3}\E[\bmX]} + \Pr\expt{\abs{(-\log \bmY - (-\log \E[\bmY])} > \frac{\varepsilon}{3}\cdot(-\log \E[\bmY])} \\
\Pr\expt{\abs{\bmz_2 - A_2} > \varepsilon A_2} &\leq \Pr\expt{\abs{\bmX - \E[\bmX]} > \frac{\varepsilon}{3}\E[\bmX]} + \Pr\expt{\abs{(-\log \bmX - (-\log \E[\bmX])} > \frac{\varepsilon}{3}\cdot (-\log \E[\bmX])}
\end{align*}
\end{lemma}
\proof{Proof.}
Note that since $\bmX, \bmY \in [0,1]$, it follows that $-\log \bmX, -\log \bmY$ are non-negative. Also, since $\bmY = 0 \implies \bmX =0$, the random variables $\bmz_1, \bmz_2$ are both well-defined (with the convention that $x\log x = 0$ when $x=0$). Further, since we also have $0 < \E[\bmX], \E[\bmY] < 1$, so that $-\log \E[\bmX] > 0$ and $-\log \E[\bmY] > 0$. The claims then follow from a straightforward application of part (ii) of lemma~\ref{lem:conc-relation}.
\Halmos \endproof

\begin{lemma}
Let $0 \leq \bmX \leq 1$ be a non-negative random variable with $1 > \E[\bmX] > 0$. Then given any $0 < \varepsilon < 1$, for all values of the random variable $\bmX$ in the interval $I := \bigg((1-\varepsilon')\E[\bmX], (1+ \varepsilon')\E[\bmX]\bigg)$, where $\varepsilon' = \frac{-\varepsilon \log \E[\bmX]}{1 - \varepsilon \log\E[\bmX]}$, we have
\[ \abs{-\log(\bmX) - (-\log\E[\bmX] )} \leq \varepsilon \cdot (-\log \E[\bmX]) \]
\label{lem:log-conc}
\end{lemma}
\proof{Proof.} Since $0 < \E[\bmX] < 1$, it means that $-\log \E[\bmX] > 0$ and consequently, $0 < \varepsilon' < 1$. In addition, it can be seen that $(1 + \varepsilon') \E[\bmX] \leq 1$ for any $0 < \varepsilon < 1$ and any $0 < \E[\bmX] < 1$, so that $I \subset [0,1]$. Consider the function $g(x) = -\log x$ and note that it is continuous and differentiable on the interval $I$. The Mean Value theorem says that given a differentiable function $g(\cdot)$ in the interval $(a,b)$, there exists $c \in (a,b)$ such that
\[ \frac{g(b) - g(a)}{b-a} = g'(c) = \frac{g(a) - g(b)}{a-b} \]
where $g'(\cdot)$ is the derivative of $g(\cdot)$. Using the mean value theorem for $g(x) = - \log x$ in the interval $I$, it follows that for all values of random variable $\bmX \in I$ there exists some $\bmz$ between $\E[\bmX]$ and $\bmX$ such that
\[ \frac{-\log \bmX - (-\log \E[\bmX])}{\bmX - \E[\bmX]} = \frac{-1}{\bmz} \]

Now since $\bmz \in I$, it follows that $\frac{1}{\bmz} \leq \frac{1}{(1 - \varepsilon')\E[\bmX]}$. Also, since $\bmX \in I$, we have $\abs{\bmX - \E[\bmX]} \leq \varepsilon' \E[\bmX]$. Then it follows:
\begin{align*}
\abs{-\log \bmX - (-\log \E[\bmX])} = \abs{\frac{-(\bmX - \E[\bmX])}{\bmz}} = \frac{\abs{\bmX - \E[\bmX]}}{\bmz}
\leq \frac{\abs{\bmX - \E[\bmX]}}{(1-\varepsilon')\E[\bmX]} \leq \frac{\varepsilon'}{1-\varepsilon'} = \varepsilon \cdot (-\log \E[\bmX])
\end{align*}
\Halmos \endproof

\section{Latent class Independent (LC-IND) model}

First, we introduce some additional notation. 
Let $\ind[A]$ denote the indicator variable taking value $1$ if an event $A$ is true and $0$ otherwise. 
Let $\xiplus$ (resp. $\ximinus$) denote the number of items rated as $+1$ (resp. $-1$) by customer $i$. In other words, $\xiplus = \sum_{j \in N(i)} \; \ind[\xij = +1]$ and $\ximinus = \sum_{j \in N(i)} \; \ind[\xij = -1]$, where recall that $N(i)$ denotes the set of items rated by customer $i$. Here, $\xij$ represents the rating provided by customer $i$ for item $j$, note that it is a random variable under the {\sf LC-IND} model. Next, let $\bm{F}_0^+ = \frac{\sum_{i'=1}^m \xipplus}{m \cdot \ell}$, so $\alphap$ is the fraction of likes (+1s) received from the customer population. 
Finally, let $\Bin(r,p)$ denote the Binomial distribution with parameters $r$ and $p$.

We begin by proving a lemma that will be used in the proof.
%

\begin{lemma}
\label{lem:pop-conc}
Consider the random variable $\bm{F}_0^+ = \frac{\sum_{i'=1}^m \xipplus}{m \cdot \ell}$. Given any $t > 0$, the following facts are true:
\begin{align*}
(i)\; \E[\alphap] = \alphapool \;\;\;\;\;\; (ii)\; \Pr\expt{\abs{\alphap - \alphapool} \geq t} \leq 2\exp\big(-2 m \ell t^2\big)
\end{align*}
\end{lemma}
\proof{Proof.}
We begin with the expectation:
\[ \E[\alphap] = \frac{\sum_{i'=1}^m \E[\xipplus]}{m \cdot \ell} = \frac{\sum_{i'=1}^m \ell\alphab_{z_i'}}{m \cdot \ell} = \frac{\sum_{k'=1}^K q_{k'} m \cdot(\ell\alphab_{k'})}{m \cdot \ell} = \sum_{k'=1}^K q_{k'} \alphab_{k'} = \alphapool \]
using the fact that proportion $q_{k'}$ of the customer population belongs to segment $k'$. 
For part $(ii)$, observe that $\bm{F}_0^+$ can be equivalently written as:
\[ \alphap = \frac{\sum \limits_{i'=1}^m \sum \limits_{j \in N(i')} \ind[\xij = +1]}{m \cdot \ell} \]
In other words, $\bm{F}_0^+$ is an average of $m \cdot \ell$ random variables, which are independent under the {\sf LC-IND} model (since each customer rates items independently). Then using Hoeffding's inequality we can show, for any $t > 0$:
\begin{equation*}
\Pr\expt{\abs{\alphap - \alphapool} \geq t} \leq 2\exp\big(-2 m \ell t^2\big)
\end{equation*}
\Halmos \endproof

\subsection{Concentration of customer projection scores}

\proof{Proof of Lemma 1.} To calculate the customer projection scores, we first need to compute the pooled estimate. Since the underlying {\sf LC-IND} model is parameterized by a single parameter that specifies the probability of liking any item, the pooled estimate is given by $\frac{\sum_{i'=1}^m \xipplus}{m \cdot \ell} = \alphap$ based on our definition earlier.
Also, let us denote the fraction of likes given by customer $i$ as $\fiplus \defas \frac{\xiplus}{\ell}$.
Then, the unidimensional projection $\projrv_i$ for customer $i$ is given by:
\begin{align*}
 \projrv_i &= \frac{-\sum_{j \in N(i)} \big(\ind[\xij = +1] \log \alphap + \ind[\xij = -1] \log (1-\alphap) \big)}
{-\sum_{j \in N(i)} \big(\alphap \log \alphap + (1-\alphap) \log (1-\alphap) \big)} \\
&= \frac{-\big(\sum_{j \in N(i)} \ind[\xij = +1]\big) \log \alphap -\big(\sum_{j \in N(i)} \ind[\xij = -1] \big) \log (1-\alphap)}
{-\big(\alphap \log \alphap + (1-\alphap) \log (1-\alphap) \big) \cdot \ell} \\
&= \frac{-\big(\frac{\xiplus}{\ell}\big) \log \alphap -\big(1 - \frac{\xiplus}{\ell}\big) \log (1- \alphap)}
{-\big( \alphap \log \alphap + (1-\alphap) \log (1-\alphap) \big)} \\
&= \frac{-\fiplus \log \alphap -\big(1 - \fiplus\big) \log (1- \alphap)}
{-\big( \alphap \log \alphap + (1-\alphap) \log (1-\alphap) \big)}
\end{align*}

Note that when $\alphap = 0$ or $1$, $\projrv_i$ is of the form $0/0$ and therefore undefined. However, we show below that with high probability, $\alphap \in \bigg(\alphapool\cdot (1-\varepsilon'), \alphapool\cdot (1+\varepsilon')\bigg)$ for some $0 < \varepsilon' < 1$ and therefore the projection score of each customer $i$ is well-defined with high probability.

{\bf Concentration of $\fiplus$}. From the generative model, it follows that the random variable representing the number of likes given by customer $i$ is a binomial random variable, i.e. $\xiplus \sim \Bin(\ell, \alphab_{z_i})$. Then, using Hoeffding's inequality we can show that for any $t > 0$:
\begin{align}
\label{eq:cust-conc}
\begin{split}
\Pr\expt{\abs{\fiplus - \alphab_{z_i}} \geq t} &\leq 2\exp\big(-2\ell t^2\big) \\
\Pr\expt{\abs{\left(1-\fiplus\right) - (1-\alphab_{z_i})} \geq t} &\leq 2\exp\big(-2\ell t^2\big)
\end{split}
\end{align}

{\bf Concentration of $-\log \alphap$ and $-\log (1-\alphap)$}. Lemma~\ref{lem:pop-conc} says that $\E[\alphap] = \alphapool \geq \alphab_{\min} > 0$ and observe that $0 \leq \alphap \leq 1$. So we can apply lemma~\ref{lem:log-conc} to the random variable $\alphap$, which says that given any $0 < \varepsilon < 1$, for all values of random variable $\alphap$ in the interval $\bigg(\alphapool\cdot (1- \varepsilon'), \alphapool\cdot (1+\varepsilon')\bigg)$ with $\varepsilon' = \frac{-\varepsilon \log \alphapool}{1 - \varepsilon \log \alphapool}$, we have: 
\begin{align}
\label{eq:logalpha}
\abs{-\log \alphap - (-\log \alphapool)} \leq \varepsilon \cdot (-\log \alphapool)
\end{align}
Now, for any $0 < \varepsilon < 1$, define $t(\varepsilon) \defas  \varepsilon \cdot \left(\frac{-\alphabpool \cdot \log \left(1-\alphabpool\right)}{1 - \log \left(1-\alphabpool\right)}\right)$, where recall that $\alphabpool = \min\set{\alphapool, 1-\alphapool}$, as defined in the statement of the lemma. Note that $\alphabpool < (1/2)$ since $\alphapool \neq (1/2)$ and therefore $t(\varepsilon)$ is well-defined. It is easy to check that $0< t(\varepsilon) \leq \alphapool \cdot \varepsilon'$ where note from above that $\varepsilon' = \frac{-\varepsilon \log \alphapool}{1 - \varepsilon \log \alphapool}$.
Then using lemma~\ref{lem:pop-conc}, we get that
\[ \Pr\expt{\abs{\alphap - \alphapool} \leq \varepsilon' \alphapool} \geq \Pr\expt{\abs{\alphap - \alphapool} \leq t(\varepsilon)} \geq 1 - 2\exp\big(-2m \ell \cdot t^2(\varepsilon) \big) \]
Then, using equation~\eqref{eq:logalpha} it follows that
\begin{equation}
\label{eq:log-conc1}
\Pr\expt{\abs{-\log \alphap - (-\log \alphapool)} \leq \varepsilon \cdot (-\log \alphapool)} \geq \Pr\expt{\abs{\alphap - \alphapool} \leq \varepsilon' \alphapool} \geq 1 - 2\exp\big(-2m\ell\cdot t^2(\varepsilon)\big)
\end{equation}
A similar sequence of arguments (using the random variable $1 - \alphap$) shows that for $\varepsilon'' = \frac{-\varepsilon \log (1-\alphapool)}{1 - \varepsilon \log (1 - \alphapool)}$ and observing that $ t(\varepsilon) \leq (1-\alphapool)\cdot \varepsilon''$:
\begin{align}
\label{eq:log-conc2}
\begin{split}
\Pr\expt{\abs{-\log (1-\alphap) -\left(-\log (1-\alphapool)\right)} \leq \varepsilon \cdot \left(-\log (1-\alphapool)\right)} &\geq \Pr\expt{\abs{\alphap - \alphapool} \leq \varepsilon''\cdot (1-\alphapool)} \\
&\geq 1 - 2\exp\big(-2m\ell\cdot t^2(\varepsilon)\big)
\end{split}
\end{align}

For ease of notation in the remainder of the proof, denote the projection score $\projrv_i = \frac{\bm{N}_{i}}{\bm{D}_{i}}$ to specify the numerator and denominator terms. 

{\bf Concentration of $\bmN_{i}$}. Let us begin with the numerator, $\bmN_{i} = -\fiplus \log \alphap - (1-\fiplus) \log (1-\alphap)$. Consider the first term: $\fiplus \cdot (-\log \alphap)$ and note that $\E[\fiplus] = \alphab_{z_i}$, $\E[\alphap] = \alphapool$. Then using lemma~\ref{lem:pen-conc} with $\bmX = \fiplus, \bmY = \alphap$ and denoting $A_1 = \hat{c}_1 \defas \alphab_{z_i}\cdot (-\log \alphapool)$:
\begin{align}
\label{eq:num1-conc}
\begin{split}
&\Pr\expt{\abs{\fiplus \cdot (-\log \alphap) - \hat{c_1}} > \varepsilon \hat{c_1}} \\
&\leq \Pr\expt{\abs{\fiplus - \alphab_{z_i}} > \frac{\varepsilon}{3}\alphab_{z_i}} + \Pr\expt{\abs{(-\log \alphap - (-\log \alphapool)} > \frac{\varepsilon}{3}\cdot(-\log \alphapool)} \\
&\leq 2\exp\bigg(-2\ell \frac{\varepsilon^2\alphab_{z_i}^2}{9}\bigg) + 2\exp\bigg(-2m\ell\cdot t^2(\varepsilon/3)\bigg) \;\;\; (\text{using equations~\eqref{eq:cust-conc} and~\eqref{eq:log-conc1}}) \\
&\leq 2\exp\bigg(-2\ell \frac{\varepsilon^2\alphab_{\min}^2}{9}\bigg) + 2\exp\bigg(-2m\ell\cdot t^2(\varepsilon/3)\bigg) \;\;\; (\text{since } \alphab_{z_i} \geq \alphab_{\min})
\end{split}
\end{align}
Similarly for the second term, observe that $\E[1 - \fiplus] = 1 - \alpha_{z_i}$, $\E[1 - \alphap] = -\log(1 - \alphapool)$. Therefore, choosing $\bmX = (1-\fiplus), \bmY = 1-\alphap$ and denoting $A_2 = \hat{c}_2 \defas (1-\alphab_{z_i})\cdot \bigg(-\log(1-\alphapool)\bigg)$ in lemma~\ref{lem:pen-conc}: 
{\small
\begin{align}
\label{eq:num2-conc}
\begin{split}
&\Pr\expt{\abs{(1-\fiplus) \cdot \bigg(-\log (1-\alphap)\bigg) - \hat{c_2}} > \varepsilon \hat{c_2}} \\
&\leq \Pr\expt{\abs{(1-\fiplus) - (1- \alpha_{z_i})} > \frac{\varepsilon}{3}\cdot(1- \alpha_{z_i})} + \Pr\expt{\abs{-\log (1-\alphap) - (-\log (1- \alphapool))} > \frac{\varepsilon}{3}\cdot (-\log (1 - \alphapool))} \\
&\leq 2\exp\bigg(-2\ell \frac{\varepsilon^2(1-\alphab_{z_i})^2}{9}\bigg) + 2\exp\bigg(-2m\cdot t^2(\varepsilon/3)\bigg) \;\;\; (\text{using equations~\eqref{eq:cust-conc} and~\eqref{eq:log-conc2}}) \\
&\leq 2\exp\bigg(-2\ell \frac{\varepsilon^2\alphab_{\min}^2}{9}\bigg) + 2\exp\bigg(-2m\cdot t^2(\varepsilon/3)\bigg) \;\;\; \left(\text{since } (1 - \alphab_{z_i}) \geq \alphab_{\min}\right)
\end{split}
\end{align}}
Combining the above two, choosing $\bmX_1 = \fiplus \cdot (-\log \alphap)$, $\bmX_2 = (1-\fiplus)\cdot (-\log (1 - \alphap))$, $a_1 = \hat{c}_1$ and $a_2 = \hat{c}_2$ in lemma~\ref{lem:conc-relation}, we get:
\begin{align}
\label{eq:num-conc}
\begin{split}
&\Pr\expt{\abs{\bmN_{i} - (\hat{c_1} + \hat{c_2})} > \frac{\varepsilon}{3} \cdot (\hat{c_1} + \hat{c_2})} \\
&\leq \Pr\expt{\abs{\fiplus\cdot (-\log \alphap) - \hat{c_1}} > \frac{\varepsilon}{3} \hat{c_1}} + \Pr\expt{\abs{(1-\fiplus)\cdot \bigg(-\log (1-\alphap)\bigg) - \hat{c_2}} > \frac{\varepsilon}{3} \hat{c_2}} \\
&\leq 4\exp\bigg(-2\ell \frac{\varepsilon^2\alphab_{\min}^2}{81}\bigg) + 4\exp\bigg(-2m\ell\cdot t^2(\varepsilon/9)\bigg) \\
&(\text{using equations~\eqref{eq:num1-conc} and~\eqref{eq:num2-conc}})
\end{split}
\end{align}

{\bf Concentration of $\bmD_{i}$}. Moving on to the denominator, $\bmD_{i} = \alphap \cdot (-\log \alphap) + (1-\alphap)\cdot (-\log (1-\alphap))$. Focusing on the first term, $\alphap \cdot (-\log \alphap)$, observe that $\E[\alphap] = \alphapool$. Again using lemma~\ref{lem:pen-conc} with $\bmX = \alphap$ and denoting $A_1 = \hat{b}_1 \defas \alphapool \cdot (-\log \alphapool)$ we get,
\begin{align*}
&\Pr\expt{\abs{\alphap \cdot (-\log \alphap) - \hat{b_1}} > \varepsilon \hat{b_1}} \\
&\leq \Pr\expt{\abs{\alphap - \alphapool} > \frac{\varepsilon}{3}\alphapool} + \Pr\expt{\abs{(-\log \alphap - (-\log \alphapool)} > \frac{\varepsilon}{3}\cdot (-\log \alphapool)} \\
&\leq 2 \exp\bigg(-2 m\ell \frac{\varepsilon^2}{9} \alphapool^2\bigg) + 2\exp\bigg(-2m\ell\cdot t^2(\varepsilon/3)\bigg) \;\;\; (\text{using lemma~\ref{lem:pop-conc} and~\eqref{eq:log-conc1}}) \\
&\leq 2 \exp\bigg(-2 m\ell \frac{\varepsilon^2}{9} \alphabpool^2\bigg) + 2\exp\bigg(-2m\ell\cdot t^2(\varepsilon/3)\bigg) \;\;\; (\text{since } \alphapool \geq \alphabpool)
\end{align*}
Similarly, for the second term choosing $\bmX = (1 - \alphap)$ and denoting $A_2 = \hat{b}_2 \defas (1-\alphapool)\cdot (-\log (1-\alphapool))$ in lemma~\ref{lem:pop-conc} we get:
{\small
\begin{align*}
&\Pr\expt{\abs{(1-\alphap)\cdot (-\log (1-\alphap)) - \hat{b_2}} > \varepsilon \hat{b_2}} \\
&\leq \Pr\expt{\abs{(1-\alphap) - (1- \alphapool)} > \frac{\varepsilon}{3}\cdot (1-\alphapool)} + \Pr\expt{\abs{(-\log (1-\alphap) - (-\log (1-\alphapool))} > \frac{\varepsilon}{3}\cdot(-\log (1-\alphapool))} \\
&= \Pr\expt{\abs{\alphap - \alphapool} > \frac{\varepsilon}{3} \cdot (1-\alphapool)} + \Pr\expt{\abs{(-\log (1-\alphap) - (-\log (1-\alphapool))} > \frac{\varepsilon}{3}\cdot (-\log (1-\alphapool))} \\
&\leq 2 \exp\bigg(-2 m\ell \frac{\varepsilon^2}{9} (1-\alphapool)^2\bigg) + 2\exp\bigg(-2m\ell\cdot t^2(\varepsilon/3)\bigg) \;\;\; (\text{using lemma~\ref{lem:pop-conc} and equation~\eqref{eq:log-conc2}}) \\
&\leq 2 \exp\bigg(-2 m\ell \frac{\varepsilon^2}{9} \alphabpool^2\bigg) + 2\exp\bigg(-2m\ell \cdot t^2(\varepsilon/3)\bigg) \;\;\; \left(\text{since } (1 -\alphapool) \geq \alphabpool\right)
\end{align*}
}

Combining the above two, choosing $\bmX_1 = \alphap \cdot (-\log \alphap)$, $\bmX_2 = (1-\alphap) \cdot (-\log (1 - \alphap))$, $a_1 = \hat{b}_1$ and $a_2 = \hat{b}_2$ in lemma~\ref{lem:conc-relation}, we get:
\begin{align}
\label{eq:den-conc}
\begin{split}
&\Pr\expt{\abs{\bmD_{i} - (\hat{b}_1 + \hat{b_2})} > \frac{\varepsilon}{3} (\hat{b}_1 + \hat{b}_2)} \\
&\leq \Pr\expt{\abs{\alphap \cdot (-\log \alphap) - \hat{b}_1} > \frac{\varepsilon}{3} \hat{b}_1} + \Pr\expt{\abs{(1-\alphap)\cdot \bigg(-\log (1-\alphap)\bigg) - \hat{b}_2} > \frac{\varepsilon}{3} \hat{b}_2} \\
&\leq 4 \exp\bigg(-2 m\ell \frac{\varepsilon^2}{81} \alphabpool^2\bigg) + 4\exp\bigg(-2m\ell\cdot t^2(\varepsilon/9)\bigg)
\end{split}
\end{align}

{\bf Concentration of $\projrv_i$}. Now that we have expressions for the concentration of the numerator and denominator, we can discuss the concentration of the projection score $\projrv_i$. Choosing $\bmX_i = \bmN_{i}$, $\bmY = \bmD_{i}$, $a_1 = \hat{c_1} + \hat{c_2}$, $b = \hat{b_1} + \hat{b_2}$ in lemma~\ref{lem:conc-relation}, we get the required concentration bound for the unidimensional projection score of customer $i$:
\begin{align*}
&\Pr\expt{\abs{\projrv_i - \frac{\hat{c_1} + \hat{c_2}}{\hat{b}_1 + \hat{b}_2}} > \varepsilon\frac{\hat{c_1} + \hat{c_2}}{\hat{b}_1 + \hat{b}_2}} \\
&=\Pr\expt{\abs{\frac{\bmN_{i}}{\bmD_{i}} - \frac{\hat{c_1} + \hat{c_2}}{\hat{b}_1 + \hat{b}_2}} > \varepsilon\frac{\hat{c_1} + \hat{c_2}}{\hat{b}_1 + \hat{b}_2}} \\
&\leq \Pr\expt{\abs{\bmN_{i} -  (\hat{c_1} + \hat{c_2})} > \frac{\varepsilon}{3}\cdot (\hat{c_1} + \hat{c_2})} +  \Pr\expt{\abs{\bmD_{i} -  (\hat{b_1} + \hat{b_2})} > \frac{\varepsilon}{3} \cdot (\hat{b_1} + \hat{b_2})} \\
&\leq 4\exp\bigg(-2\ell \frac{\varepsilon^2 \alphab_{\min}^2}{81}\bigg) + 4 \exp\bigg(-2 m\ell \frac{\varepsilon^2}{81} \alphabpool^2\bigg) + 8\exp\bigg(-2m\ell\cdot t^2(\varepsilon/9)\bigg) \\
&\left(\text{from equations} ~\eqref{eq:num-conc} \text{ and }~\eqref{eq:den-conc} \right) \\
&\leq 4\exp\bigg(-2\ell \frac{\varepsilon^2 \alphab_{\min}^2}{81}\bigg) + 12\exp\bigg(-2m\ell\cdot t^2(\varepsilon/9)\bigg) \\
&\left(\text{since } \frac{\log^2 (1-\alphabpool)}{(1- \log(1-\alphabpool))^2} < 1 \right)
\end{align*}
Finally, note that $\hat{c_1} + \hat{c_2} = H(\alphab_{z_i}, \alphapool)$, the cross-entropy between the distributions $\Ber(\alphab_{z_i})$ and $\Ber(\alphapool)$ and; $\hat{b_1} + \hat{b_2} = H(\alphapool)$, the binary entropy function at $\alphapool$. In other words, the unidimensional projection of customer $i$ in segment $k$ concentrates around the ratio $\frac{H(\alphak, \alphapool)}{H(\alphapool)}$ with high probability, as $\ell \to \infty$.
\Halmos \endproof

\proof{Proof of Theorem 1.} The result follows directly from the concentration of the projection score to the ratio $\frac{H(\alphak, \alphapool)}{H(\alphapool)}$, refer to the discussion after Lemma 1 in the main text.
\endproof

\subsection{Asymptotic recovery of true segments}
Having established the concentration of the customer projection scores, we next discuss the error-rate of classification into the underlying segments. We begin by proving some useful lemmas. All notations are as stated in the main text, unless otherwise introduced.

\begin{lemma}
\label{lem:nn-dist}
Let $k_1, k_2$ be two arbitrary segments. Then for customer $i$, we have 
\[ \frac{\abs{\projrv_i - H_{k_1}}}{H_{k_1}} \leq \frac{\abs{H_{k_1} - H_{k_2}}}{2 \cdot \max(H_{k_1}, H_{k_2})} \implies \frac{\abs{\projrv_i - H_{k_1}}}{H_{k_1}} \leq \frac{\abs{\projrv_i - H_{k_2}}}{H_{k_2}} \]
\end{lemma}
\proof{Proof.}
Consider the following:
\begin{align*}
\frac{\abs{H_{k_1} - H_{k_2}}}{\max(H_{k_1}, H_{k_2})} &= \frac{\abs{(H_{k_1} - \projrv_i) + (\projrv_i - H_{k_2})}}{\max(H_{k_1}, H_{k_2})}\\
&\leq \frac{\abs{\projrv_i - H_{k_1}}}{\max(H_{k_1}, H_{k_2})} + \frac{\abs{\projrv_i - H_{k_2}}}{\max(H_{k_1}, H_{k_2})} \\
&(\text{using triangle inequality}) \\
&\leq \frac{\abs{\projrv_i - H_{k_1}}}{H_{k_1}} + \frac{\abs{\projrv_i - H_{k_2}}}{H_{k_2}} \\
&\leq \frac{\abs{H_{k_1} - H_{k_2}}}{2 \cdot \max(H_{k_1}, H_{k_2})} + \frac{\abs{\projrv_i - H_{k_2}}}{H_{k_2}} \\
&(\text{follows from the hypothesis of the lemma}) \\
\end{align*}
Therefore we have that 
\[ \frac{\abs{\projrv_i - H_{k_2}}}{H_{k_2}} \geq \frac{\abs{H_{k_1} - H_{k_2}}}{2 \cdot \max(H_{k_1}, H_{k_2})} \geq \frac{\abs{\projrv_i - H_{k_1}}}{H_{k_1}} \]
\Halmos \endproof

\begin{lemma}
\label{lem:lc_ind_const}
Consider the constant $\Lambda$ defined in Theorem 2:
\[ \Lambda = \frac{\abs{\log \frac{\alphapool}{1-\alphapool}} \min_{k = 1,2,\dotsb, K} (\alpha_{k+1} - \alpha_{k}) }{2 \abs{\log \amin}} \]
Then, it follows that $\Lambda \leq \min_{k \neq k'} \frac{\abs{H_{k} - H_{k'}}}{2\cdot \max(H_{k}, H_{k'})} < 1$.
\end{lemma}
\proof{Proof.}
Recall that $H_k = \frac{H(\alpha_k, \alphapool)}{H(\alphapool)}$. Therefore, for any two segments $k \neq k'$, we have:
\[ \Lambda_{kk'} \defas \frac{\abs{H_{k} - H_{k'}}}{2\cdot \max(H_{k}, H_{k'})} = \frac{\abs{H(\alpha_{k}, \alphapool) - H(\alpha_{k'}, \alphapool)}}{2\cdot \max\set{H(\alpha_k, \alphapool), H(\alpha_{k'}, \alphapool)}} \]
Next, observe that $H(\alpha_k, \alphapool) = -\alpha_k \log\frac{\alphapool}{1-\alphapool} - \log (1-\alphapool)$, so that
\[ \abs{H(\alpha_{k}, \alphapool) - H(\alpha_{k'}, \alphapool)} = \abs{\log \frac{\alphapool}{1-\alphapool}} \abs{\alpha_{k} - \alpha_{k'}} \]
Now suppose $\alphapool > \frac{1}{2}$, this means that $H(\alpha_k, \alphapool)$ is decreasing with $\alpha_k$ so that we have
\begin{equation*}
\max\set{H(\alpha_k, \alphapool), H(\alpha_{k'}, \alphapool)} \leq H(\alpha_{\min}, \alphapool) \leq H(\alpha_{\min}, 1-\alpha_{\min}) \leq -\log \amin
\end{equation*}
where the second inequality follows from the fact that $\alphapool \leq 1 - \alpha_{\min}$ and the last inequality from the fact that $-\log (1-\amin) \leq -\log \amin$. Similarly, when $\alphapool < \frac{1}{2}$, we have
\[ \max\set{H(\alpha_k, \alphapool), H(\alpha_{k'}, \alphapool)} \leq H(1 - \alpha_{\min}, \alphapool) \leq  H(1 - \alpha_{\min}, \alpha_{\min}) = H(\amin, 1-\amin)  \leq -\log \amin\]
where the second inequality is true because $\alphapool \geq \amin$. 

Combining the above observations and using the fact that $\abs{\alpha_k - \alpha_{k'}} \geq \min_{k''=1,2,\ldots,K-1} (\alpha_{k''+1} - \alpha_{k''})$ for all $k \neq k'$, we get $\Lambda_{kk'} \geq \Lambda$ for all $k \neq k'$. Further, observe that $\Lambda_{kk'} < 1$ because $H_k > 0$ for all $1 \leq k \leq K$. Therefore, $\Lambda \leq \min_{k\neq k'} \Lambda_{kk'} < 1$.
\Halmos \endproof

\begin{lemma}
\label{lem:misclass}
Consider customer $i$ and suppose we have the following:
\[ \frac{\abs{\projrv_i - H_{z_i}}}{H_{z_i}} \leq \frac{\abs{H_{z_i} - H_{k'}}}{2\cdot \max(H_{z_i}, H_{k'})} \;\;\; \forall \; k' \neq z_i\]
Then it follows that $\hbmz(i) = z_i$, i.e we correctly classify customer $i$. Conversely, we have
\[ \Pr\expt{\hbmz(i) \neq z_i} \leq \Pr\expt{\abs{\projrv_i - H_{z_i}} > \Lambda \cdot H_{z_i}} \]
\end{lemma}
\proof{Proof.}
Using lemma~\ref{lem:nn-dist} we obtain that $\frac{\abs{\projrv_i - H_{z_i}}}{H_{z_i}} \leq \frac{\abs{\projrv_i - H_{k'}}}{H_{k'}}$ for all $k' \neq z_i$. This means that $\argmin_{k \in [K]} \frac{\abs{\projrv_i - H_k}}{H_k} = z_i$. For the second part of the claim, observe that if $\hbmz(i) \neq z_i$ then there exists some $k \neq z_i$ such that $\frac{\abs{\projrv_i - H_{z_i}}}{H_{z_i}} > \frac{\abs{H_{z_i} - H_{k}}}{2\cdot \max(H_{z_i}, H_{k})} \geq \Lambda$, which follows from Lemma~\ref{lem:lc_ind_const} above. In other words,
\[ \hbmz(i) \neq z_i \implies \abs{\projrv_i - H_{z_i}} > \Lambda \cdot H_{z_i} \] and the claim follows.
\Halmos \endproof

\proof{Proof of Theorem 2.} 
The probability that customer $i$ is misclassified by the nearest-neighbor classifier $\hbmz(\cdot)$ is given by:
\begin{align*}
\Pr\expt{\hbmz(i) \neq z_i} &\leq \Pr\expt{\abs{\projrv_i - H_{z_i}} >  \Lambda \cdot H_{z_i}} \;\;\; (\text{using lemma~\ref{lem:misclass}}) \\
&\leq 4\exp\bigg(-2 \ell \frac{\Lambda^2 \alphab_{\min}^2}{81}\bigg) + 12\exp\left( \frac{-2 m\cdot \ell \cdot  \Lambda^2 \alphabpool^2 \log^2 (1-\alphabpool)}{81 \big(1- \log (1-\alphabpool)\big)^2}\right) \\
&(\text{using result of Lemma 1})
\end{align*}

Now given some $0 < \delta < 1$, suppose that the number of observations from each customer satisfy:
\[ \ell \geq \frac{648}{\lambda^2} \cdot \left(\frac{\log \alpha_{\min}}{\log (1-\amin) \cdot \amin}\right)^2 \cdot \frac{1}{\log^2\frac{\alphapool}{1-\alphapool}} \cdot \log (16/\delta) \]
Then, it follows from above that
\begin{align*}
\Pr\expt{\hbmz(i) \neq z_i} &\leq 4\exp\bigg(-2\ell \frac{\Lambda^2 \alphab_{\min}^2}{81}\bigg) + 12\exp\left( \frac{-2 m\cdot \ell \cdot \Lambda^2 \alphabpool^2 \log^2 (1-\alphabpool)}{81 \big(1- \log (1-\alphabpool)\big)^2}\right) \\
&\leq 4\exp\bigg(-2\ell \frac{\Lambda^2 \alphab_{\min}^2}{81}\bigg) + 12\exp\left( \frac{-2 m\cdot \ell \cdot \Lambda^2 \amin^2 \log^2 (1-\amin)}{81 \big(1- \log (1-\amin)\big)^2}\right) \\
&(\text{since }\alphabpool \geq \amin) \\
&\leq 4\exp\bigg(-2\ell \frac{\Lambda^2 \alphab_{\min}^2 \log^2 (1-\amin)}{81 \left(1 - \log (1-\amin)\right)^2}\bigg) + 12\exp\left( \frac{-2 m\cdot \ell \cdot \Lambda^2\amin^2 \log^2 (1-\amin)}{81 \big(1- \log (1-\amin)\big)^2}\right) \\
&\bigg(\text{since } \log^2 (1- \amin) < \left(1 - \log (1-\amin)\right)^2\bigg) \\
&\leq 16\exp\bigg(-2\ell \frac{\Lambda^2 \alphab_{\min}^2 \log^2 (1-\amin)}{81 \left(1 - \log (1-\amin)\right)^2}\bigg) \\
&(\text{since } m \geq 1) \\
&= 16\exp\bigg(-\ell \frac{\lambda^2 \alphab_{\min}^2 \log^2  \frac{\alphapool}{1-\alphapool} \log^2 (1-\amin)}{162 \cdot \log^2 \amin \cdot \left(1 - \log (1-\amin)\right)^2}\bigg) \\
&(\text{substituting the value of } \Lambda) \\
&\leq 16\exp\bigg(-\ell \frac{\lambda^2 \alphab_{\min}^2 \log^2  \frac{\alphapool}{1-\alphapool} \log^2 (1-\amin)}{648 \cdot \log^2 \amin}\bigg) \\
& \bigg(\text{since } \amin < \frac{1}{2} \implies 1 - \log (1-\amin) < 2 \bigg) \\
&\leq \delta \\
& \bigg(\text{using the bound on } \ell \bigg)
\end{align*}

Next, suppose that $m \cdot \frac{\log^2 (1-\alphabpool)}{\big(1- \log (1-\alphabpool)\big)^2} \geq 1$, and observe that $\alphabpool \geq \alpha_{\min}$. Then we get,
\begin{align*}
\Pr\expt{\hbmz(i) \neq z_i} &\leq \Pr\expt{\abs{\projrv_i - H_{z_i}} >  \Lambda \cdot H_{z_i}} \;\;\; (\text{using lemma~\ref{lem:misclass}}) \\
&\leq 4\exp\bigg(-2\ell \frac{\Lambda^2\alphab_{\min}^2}{81}\bigg) + 12\exp\left( \frac{-2 m\cdot \ell \cdot  \Lambda^2 \alphabpool^2 \log^2 (1-\alphabpool)}{81 \big(1- \log (1-\alphabpool)\big)^2}\right) \\
&(\text{using result of Lemma 1}) \\
&\leq 16\exp\bigg(-2\ell \frac{\Lambda^2\alphab_{\min}^2}{81}\bigg)
\end{align*}
Substituting $\ell = \log n$ we get the desired result.

\section{Latent class Independent Category (LC-IND-CAT) model}
Recall that segment $k$ is characterized by $B$-dimensional vector $\alphavec_k$ such that $\alphab_{kb}$ represents the probability of liking any item $j \in \I_b$. 
Let $\xibplus$ denote the number of likes given by customer $i$ for items in category $b$, i.e. $\xibplus = \sum_{j \in N_b(i)} \ind[\xij = +1]$, where $N_b(i) \subset \I_b$ denotes the collection of items of category $b$ rated by customer $i$. To calculate the customer projection scores, we first need to compute the pooled distribution. Since the underlying {\sf LC-IND-CAT} model is parameterized by a vector of length $B$ for each segment, the pooled estimate is given by $\bar{\bm{F}}_0^+ = (\bm{F}_{01}^+, \bm{F}_{02}^+, \ldots, \bm{F}_{0B}^+)$ where:
\[ \fbplus \defas \frac{\sum_{i'=1}^m \xibpplus}{m \cdot \ell_b} \;\;\; \forall \; b \in [B] \]
where recall that $\ell_b$ is the number of items in category $b$ that each customer rates. Also, let us denote the fraction of likes given by customer $i$ for category $b$ items as $\fibplus \defas \frac{\xibplus}{\ell_b}$. We first prove a lemma that will be useful in the proof:
\begin{lemma}
\label{lem:pop-cat-conc}
Given any $t > 0$, for each category $b \in [B]$, the following facts are true:
\begin{align*}
{\rm (i)} & \;\; \xibplus \sim \Bin(\ell_b, \alpha_{z_ib}) \\
{\rm (ii)} & \;\; \E[\fbplus] = \alphacatpool{b} \\
{\rm (iii)} & \;\;\Pr\expt{\abs{\fbplus - \alphacatpool{b}} \geq t} \leq 2\exp\big(-2 m\cdot \ell_b\cdot t^2\big)
\end{align*}
\end{lemma}
\proof{Proof.}
Lets begin with part (i). Observe that $\xibplus = \sum\limits_{j \in N_b(i)} \ind[\xij = +1]$. Based on the generative model, we have that $\ind[\xij = +1]$ are independent and identically distributed such that $\Pr[\xij = +1] = \alpha_{z_ib}$. The claim then follows.

For part (ii) observe that,
\[ \E[\fbplus] = \frac{\sum_{i'=1}^m \E[\xibpplus]}{m \cdot \ell_b} = \frac{\sum_{i'=1}^m \ell_b\alphab_{z_{i'}b}}{m \cdot \ell_b} = \frac{\sum_{k'=1}^K (q_{k'} m) \cdot(\ell_b\alphab_{k'b})}{m \cdot \ell_b} = \sum_{k'=1}^K q_{k'} \alphab_{k'b} = \alphacatpool{b} \]
using the fact that proportion $q_{k'}$ of the customer population belongs to segment $k'$.
For part (iii), observe that $\fbplus$ can be written as:
\[ \fbplus = \frac{\sum \limits_{i'=1}^m \sum_{j \in N_b(i)} \ind[\xij = +1]} {m \cdot \ell_b} \]
In other words, $\fbplus$ is an average of $m \cdot \ell_b$ random variables, which are independent under the {\sf LC-IND-CAT} model (since ratings for items within the same category are independent and the observations of different customers are generated independently). Then using Hoeffding's inequality we can show, for any $t > 0$:
\begin{equation*}
\Pr\expt{\abs{\fbplus - \alphacatpool{b}} \geq t} \leq 2\exp\big(-2 m \cdot \ell_b \cdot t^2\big)
\end{equation*}

\Halmos \endproof

\subsection{Concentration of customer projection score vectors}

{\em Proof of Lemma 2.} For customer $i$ that belongs to segment $k$, the projection score computed by our algorithm, $\projrvec_i = (\projrv_{i1}, \projrv_{i2}, \dotsb, \projrv_{iB})$ is a $B$-dimensional vector, where: 
\begin{align*}
 \projrv_{ib} &= \frac{-\sum_{j \in N_b(i)} \big(\ind[\xij = +1] \log \fbplus + \ind[\xij = -1] \log (1-\fbplus) \big)}
{-\sum_{j \in N_b(i)} \big(\fbplus \log \fbplus + (1-\fbplus) \log (1-\fbplus) \big)} \\
&= \frac{-\big(\sum_{j \in N_b(i)} \ind[\xij = +1]\big) \log \fbplus -\big(\sum_{j \in N_b(i)} \ind[\xij = -1] \big) \log (1-\fbplus)}
{-\big(\fbplus \log \fbplus + (1-\fbplus) \log (1-\fbplus) \big) \cdot \ell_b} \\
&= \frac{-\big(\frac{\xibplus}{\ell_b}\big) \log \fbplus -\big(1 - \frac{\xibplus}{\ell_b}\big) \log (1- \fbplus)}
{-\big(\fbplus \log \fbplus + (1-\fbplus) \log (1-\fbplus) \big)} \\
&= \frac{-\fibplus \log \fbplus -\big(1 - \fibplus\big) \log (1- \fbplus)}
{-\big(\fbplus \log \fbplus + (1-\fbplus) \log (1-\fbplus) \big)}
\end{align*}

Observe that the precise sequence of arguments given in the proof of Lemma 1 earlier can be repeated, for each item category $b$ separately. More precisely, it follows that, given any $0 < \varepsilon < 1$, and for each $b \in [B]$:
\[ \Pr\expt{\abs{\projrv_{ib} - \frac{H(\alphab_{z_ib}, \alphacatpool{b})}{H(\alphacatpool{b})}} > \varepsilon\frac{H(\alphab_{z_ib}, \alphacatpool{b})}{H(\alphacatpool{b})}} \leq 4\exp\bigg(-2\ell_b \frac{\varepsilon^2 \alphab_{\min}^2}{81}\bigg) + 12\exp\bigg(-2m \cdot\ell_b \cdot t_b^2(\varepsilon/9)\bigg) \]
where $t_b(\varepsilon) \defas \varepsilon \cdot \left(\frac{-\baralphabpool \log (1 - \baralphabpool)}{1 - \log(1-\baralphabpool)}\right)$ and $\baralphabpool = \min\set{\alphacatpool{b}, 1 - \alphacatpool{b}}$.

Then, we consider the convergence of the vector $\projrvec_i$. Define the $B$-dimensional vector $\bm{H}_k = (H_{k1}, H_{k2}, \dotsb, H_{kB})$ for each $k \in [K]$ such that $H_{kb} = \frac{H(\alphab_{kb}, \alphacatpool{b})}{H(\alphacatpool{b})}$, note that each $H_{kb} > 0$ (since all parameters are bounded), so that $\norm{\bmH_k}_1 > 0$ for all $k \in [K]$. Then, using lemma~\ref{lem:conc-relation}(iv) it follows that:
\begin{align*}
\Pr\expt{\norm{\projrvec_i - \bm{H}_{z_i}}_1 > \varepsilon \norm{\bm{H}_{z_i}}_1} &= \Pr\expt{\sum\limits_{b=1}^B \abs{\projrv_{ib} - H_{z_ib}} > \varepsilon \cdot \left(\sum\limits_{b=1}^B H_{z_ib}\right)} \\
&\leq \sum \limits_{b=1}^B \Pr\expt{\abs{\projrv_{ib} - H_{z_ib}} > \varepsilon H_{z_ib}} \\
&\leq \sum \limits_{b=1}^B 4\exp\bigg(-2\ell_b \frac{\varepsilon^2 \alphab_{\min}^2}{81}\bigg) + 12\exp\bigg(-2m \cdot \ell_b \cdot t_b^2(\varepsilon/9)\bigg)  \\
&\leq 4\cdot B \cdot \exp\bigg(-2\ell_{\min} \frac{\varepsilon^2 \alphab_{\min}^2}{81}\bigg) + 12\cdot B\cdot \exp\bigg(-2m \cdot \ell_{\min} \cdot t_{\min}^2(\varepsilon/9)\bigg)
\end{align*}
where $t_{\min}(\varepsilon) \defas \varepsilon \cdot \left(\frac{-\hatalphapool \log (1 - \hatalphapool)}{1 - \log(1-\hatalphapool)}\right)$ and $\hatalphapool = \min_{b \in [B]} \baralphabpool$. The last inequality follows from the facts that $\ell_b \geq \ell_{\min}$ and $\baralphabpool \geq \hatalphapool$ for all $b \in [B]$. Substituting for $t_{\min}(\varepsilon)$ in the equation above establishes the result.
\Halmos \endproof
\subsection{Asymptotic recovery of customer segments}
We begin with proving analogous versions of Lemmas A5-A7.
\begin{lemma}
\label{lem:nn-dist-cat}
Let $k_1, k_2$ be two arbitrary segments and $\norm{\cdot}$ be an arbitrary norm on $\Real^B$. Then for customer $i$, we have 
\[ \frac{\norm{\projrvec_i - \bmH_{k_1}}}{\norm{\bmH_{k_1}}} \leq \frac{\norm{\bmH_{k_1} - \bmH_{k_2}}}{2 \cdot \max(\norm{\bmH_{k_1}}, \norm{\bmH_{k_2}})} \implies \frac{\norm{\projrvec_i - \bmH_{k_1}}}{\norm{\bmH_{k_1}}} \leq \frac{\norm{\projrvec_i - \bmH_{k_2}}}{\norm{\bmH_{k_2}}} \]
\end{lemma}
\proof{Proof.}
The proof follows from a similar argument as in Lemma~\ref{lem:nn-dist}.
\Halmos \endproof

\begin{lemma}
\label{lem:lc_ind_cat_const}
Consider the constant $\catcons$ defined in Theorem 4. Then it follows that $\catcons \leq \min_{k' \neq k} \frac{\norm{\bmH_{k} - \bmH_{k'}}}{2\cdot \max\left(\norm{\bmH_{k}}, \norm{\bmH_{k'}}\right)} < 1$.
\end{lemma}
\proof{Proof.}
Recall that $\bmH_k \in \Real^B$ such that $H_{kb} = \frac{H(\alpha_{kb}, \alphacatpool{b})}{H(\alphacatpool{b})}$ for each $1 \leq b \leq B$. Therefore, we can write
\[ \catcons_{kk'} \defas \frac{\norm{\bmH_{k} - \bmH_{k'}}}{2\cdot \max(\norm{\bmH_{k}}, \norm{\bmH_{k'}})} = \frac{\sum_{b=1}^B \frac{\abs{H(\alpha_{kb}, \alphacatpool{b}) - H(\alpha_{k'b}, \alphacatpool{b})}}{H(\alphacatpool{b})}}{2\cdot \max(\norm{\bmH_{k}}, \norm{\bmH_{k'}}}) \]
Next, observe that $H(\alpha_{kb}, \alphacatpool{b}) = -\alpha_{kb} \log\frac{\alphacatpool{b}}{1-\alphacatpool{b}} - \log (1-\alphacatpool{b})$, so that
\[ \abs{H(\alpha_{kb}, \alphacatpool{b}) - H(\alpha_{k'b}, \alphacatpool{b})} = \abs{\log \frac{\alphacatpool{b}}{1-\alphacatpool{b}}} \abs{\alpha_{kb} - \alpha_{k'b}} \]
Note that $H(\alphacatpool{b}) \leq 1$ for any category $b$, using the definition of the binary entropy function. Therefore it follows,
\begin{align*}
\sum_{b=1}^B \frac{\abs{H(\alpha_{kb}, \alphacatpool{b}) - H(\alpha_{k'b}, \alphacatpool{b})}}{H(\alphacatpool{b})} &= \sum_{b=1}^B \frac{\abs{\log \frac{\alphacatpool{b}}{1-\alphacatpool{b}}} \abs{\alpha_{kb} - \alpha_{k'b}}}{H(\alphacatpool{b})} \\
&\geq \sum_{b=1}^B \abs{\log \frac{\alphacatpool{b}}{1-\alphacatpool{b}}} \abs{\alpha_{kb} - \alpha_{k'b}} \\
&\geq \gamma
\end{align*}
where $\gamma$ is as defined in the theorem. Next, consider $\norm{\bmH_k}$ for some segment $k$:
\[ \norm{\bmH_k} = \sum_{b=1}^B \frac{H(\alpha_{kb}, \alphacatpool{b})}{H(\alphacatpool{b})} \leq \frac{1}{H_{\min}}\sum_{b=1}^B H(\alpha_{kb}, \alphacatpool{b}) \]
where $H_{\min} \defas H(\amin)$. The above statement is true because $\amin \leq \alphacatpool{b} \leq 1 -\amin$ and the binary entropy function is symmetric around $\frac{1}{2}$ so that $H(\amin) = H(1-\amin)$, from which it follows $H(\alphacatpool{b}) \geq H_{\min}$ for any category $b$.
Further since $\alpha_{\min} \leq \alpha_{kb} \leq 1 - \alpha_{\min}$, we can use the argument from lemma~\ref{lem:lc_ind_const} to obtain $H(\alpha_{kb}, \alphacatpool{b}) \leq \abs{\log \amin}$ for all segments $k$ and item categories $b$. This further implies that $\norm{\bmH_k} \leq \frac{B \cdot \abs{\log \amin}}{H_{\min}}$ for all segments $k \in [K]$. Finally observe that $H_{\min} = -\amin\log \amin -(1-\amin) \log (1-\amin) \geq -\log (1-\amin)$, so that $\norm{\bmH_k} \leq \frac{B \cdot \abs{\log \amin}}{H_{\min}} \leq \frac{B \cdot \abs{\log \amin}}{\abs{\log (1-\amin)}}$.

Combining the above observations, we get that $\catcons_{kk'} \geq \catcons$ for all $k \neq k'$. Further, observe that $\catcons_{kk'} < 1$ since the vectors $\bmH_k$ contain only non-negative entries. Therefore, $\catcons \leq \min_{k \neq k'} \catcons_{kk'} < 1$.
\Halmos \endproof

\begin{lemma}
\label{lem:misclass_cat}
Consider a customer $i$ and suppose the following is true for an arbitrary choice of norm $\norm{\cdot}$ on $\Real^B$:
\[ \frac{\norm{\projrvec_i - \bmH_{z_i}}}{\norm{\bmH_{z_i}}} \leq \frac{\norm{\bmH_{z_i} - \bmH_{k'}}}{2\cdot \max(\norm{\bmH_{z_i}}, \norm{\bmH_{k'}})} \;\;\; \forall \; k' \neq z_i\]
Then it follows that $\hbmz_2(i) = z_i$, i.e we correctly classify customer $i$. Conversely, we have
\[ \Pr\expt{\hbmz_2(i) \neq z_i} \leq \Pr\expt{\norm{\projrvec_i - \bmH_{z_i}} > \Gamma \cdot \norm{\bmH_{z_i}}} \]
\end{lemma}
\proof{Proof.}
The proof follows from an identical argument as in Lemma~\ref{lem:misclass}, using the results of Lemmas A9 and A10 above.
\Halmos \endproof

\vspace{0.2in}
\proof{Proof of Theorem 4.} The probability that a customer $i$ is misclassified by the nearest-neighbor classifier $\hbmz_2(\cdot)$ is given by:
\begin{align*}
\Pr\expt{\hbmz_2(i) \neq z_i} &\leq \Pr\expt{\norm{\projrvec_i - H_{z_i}}_1 >  \Gamma \cdot \norm{\bmH_{z_i}}_1} \;\;\; (\text{using lemma~\ref{lem:misclass_cat}}) \\
&\leq 4\cdot B \cdot \exp\bigg(-2\ell_{\min} \frac{\Gamma^2 \alphab_{\min}^2}{81}\bigg) + 12\cdot B\cdot \exp\left( \frac{-2 m\cdot \ell_{\min} \cdot \Gamma^2 \hatalphapool^2 \log^2 (1-\hatalphapool)}{81 \big(1- \log (1-\hatalphapool)\big)^2}\right) \\
&(\text{follows from Lemma 2})
\end{align*}

Now given some $0 < \delta < 1$, suppose that the number of observations from each customer satisfy:
\[ \ell_{\min} \geq \frac{648B^2}{\gamma^2} \cdot \left(\frac{\log \alpha_{\min}}{\log^2 (1-\amin) \cdot \amin}\right)^2 \log (16B/\delta)
\]
Then, it follows from above that
\begin{align*}
\Pr\expt{\hbmz_2(i) \neq z_i} 
&\leq 4\cdot B\cdot \exp\bigg(-2\ell_{\min} \frac{\Gamma^2 \alphab_{\min}^2}{81}\bigg) + 12\cdot B \cdot \exp\left( \frac{-2 m \cdot \ell_{\min} \cdot \Gamma^2 \cdot \hatalphapool^2 \log^2 (1-\hatalphapool)}{81 \big(1- \log (1-\hatalphapool)\big)^2}\right) \\
&\leq 4\cdot B \cdot \exp\bigg(-2\ell_{\min} \frac{\Gamma^2 \alphab_{\min}^2}{81}\bigg) + 12\cdot B \cdot \exp\left( \frac{-2 m \cdot \ell_{\min} \cdot \Gamma^2 \cdot \amin^2 \log^2 (1-\amin)}{81 \big(1- \log (1-\amin)\big)^2}\right) \\
&(\text{since }\hatalphapool \geq \amin) \\
&\leq 4\cdot B\cdot \exp\bigg(-2\ell_{\min} \frac{\Gamma^2 \alphab_{\min}^2 \log^2 (1-\amin)}{81 \left(1 - \log (1-\amin)\right)^2}\bigg) + 12\cdot B \cdot \exp\left(\frac{-2m \cdot \ell_{\min} \cdot \Gamma^2\amin^2 \log^2 (1-\amin)}{81 \big(1- \log (1-\amin)\big)^2}\right) \\
&\bigg(\text{since } \log^2 (1- \amin) < \left(1 - \log (1-\amin)\right)^2\bigg) \\
&\leq 16 \cdot B \cdot \exp\bigg(-2\ell_{\min} \frac{\Gamma^2 \alphab_{\min}^2 \log^2 (1-\amin)}{81 \left(1 - \log (1-\amin)\right)^2}\bigg) \\
&(\text{since } m \geq 1) \\
&= 16\cdot B \cdot \exp\bigg(-2\ell_{\min} \frac{\gamma^2 \cdot \log^2 (1 - \amin) \cdot \alphab_{\min}^2 \log^2 (1-\amin)}{324 B^2 \cdot \log^2 \amin \cdot \left(1 - \log (1-\amin)\right)^2}\bigg) \\
&(\text{substituting value of } \Gamma) \\
&\leq 16\cdot B \cdot \exp\bigg(-\ell_{\min} \frac{\gamma^2 \cdot \log^2 (1 - \amin) \cdot \alphab_{\min}^2 \log^2 (1-\amin)}{648 B^2 \cdot \log^2 \amin}\bigg) \\
&\bigg(\text{since } \amin < \frac{1}{2} \implies 1 - \log (1-\amin) < 2 \bigg)\\
&\leq \delta \\
& \bigg(\text{using the bound on } \ell_{\min} \bigg)
\end{align*}

Finally, suppose that $m \cdot \frac{\log^2 (1-\hatalphapool)}{\big(1- \log (1-\hatalphapool)\big)^2} \geq 1$, and observe that $\hatalphapool \geq \alpha_{\min}$. Then we get,
\begin{align*}
\Pr\expt{\hbmz_2(i) \neq z_i} &\leq \Pr\expt{\norm{\projrvec_i - \bmH_{z_i}}_1 >  \Gamma \cdot \norm{\bmH_{z_i}}_1} \;\;\; (\text{using lemma~\ref{lem:misclass_cat}}) \\
&\leq 4\cdot B\cdot \exp\bigg(-2\ell_{\min} \frac{\Gamma^2 \alphab_{\min}^2}{81}\bigg) + 12\cdot B \cdot \exp\left( \frac{-2 m \cdot \ell_{\min} \cdot \Gamma^2 \cdot \hatalphapool^2 \log^2 (1-\hatalphapool)}{81 \big(1- \log (1-\hatalphapool)\big)^2}\right) \\
&(\text{follows from Lemma 2}) \\
&\leq 4\cdot B \cdot \exp\bigg(-2\ell_{\min} \frac{\Gamma^2 \alphab_{\min}^2}{81}\bigg) + 12\cdot B \cdot \exp\left( -2\ell_{\min} \frac{\Gamma^2 \alphab_{\min}^2}{81}\right) \\
&= 16\cdot B \cdot \exp\bigg(-2\ell_{\min} \frac{\Gamma^2 \alphab_{\min}^2}{81}\bigg)
\end{align*}

Substituting $\ell_{\min} = \log n$ we get the desired result.

\section{EM Algorithm for Latent Class Segmentation}
\label{app:lc-em}
Let $\bTheta = \expt{q_1, q_2,\dotsb, q_K, \alpha_1,\alpha_2, \dotsb, \alpha_K}$ denote the set of all parameters (refer to the setup in section~\ref{sec:simul_exp}). The total number of parameters is therefore $K + K = 2\cdot K$. For ease of notation, we assume that the customer-item preference graph is complete but the EM algorithm can be immediately extended for the case of incomplete graphs. Let $\calD = \set{\bmx_1, \bmx_2, \ldots, \bmx_m}$ be the observed rating vectors from the $m$ customers. Then assuming that the vectors $\bmx_i$ are sampled IID from the population mixture distribution, the log-likelihood of the data can be written as:
\begin{equation}
\label{eq:mle-em}
\log \Pr[\calD |\bTheta] = \sum_{i=1}^{m} \log \sum_{k=1}^K q_k \left(\prod_{j=1}^n \alpha_{k}^{\ind[x_{ij} = +1]} {(1-\alpha_{k})}^{\ind[x_{ij} = -1]}\right)
\end{equation}
where $\ind[\cdot]$ denotes the indicator function. The MLE for the parameters can be computed via the EM algorithm by introducing the latent variables corresponding to the true segment of each customer, which we denote by $\bm{z} = [z_1,z_2,\ldots,z_m]$ where $z_i \in [K]$ denotes the true segment of customer $i$. The complete log-likelihood can then be written as

\[ \log \Pr[\calD, \bm{z} |\bTheta] = \sum_{i=1}^{m} \sum_{k=1}^K \ind[z_i = k] \log \left(q_k \prod_{j=1}^n \alpha_{k}^{\ind[x_{ij} = +1]} {(1-\alpha_{k})}^{\ind[x_{ij} = -1]}\right) \] 

The EM algorithm executes the following two steps in each iteration:
\begin{itemize}
\item {\bf E-step}: Given the data $\calD$ and the current estimate of the parameters $\bTheta^{(t)}$, we compute the conditional expectation of the log-likelihood (w.r.t to the unknown customer segments $\bm{z}$) as
\[ \mathbb{E} \set{\log \Pr[\calD, \bm{z} |\bTheta^{(t)}]} = \sum_{i=1}^m \sum_{k=1}^K \gamma_{ik}^{(t)} \set{\log q_k^{(t)} + \sum_{j=1}^n \left({\ind[x_{ij} = 1]} \log \alpha_{k}^{(t)} + \ind[x_{ij} = -1] \log (1-\alpha_{k}^{(t)})\right)} \] 
Here $\gamma_{ik}^{(t)} = \Pr[z_{i} = k \; | \; \calD, \bTheta^{(t)}]$ is the posterior probability of customer $i$'s latent segment being equal to $k \in [K]$, conditioned on the observed ratings and the current model parameters. We can compute $\gamma_{ik}^{(t)}$ using Bayes theorem as follows
\[ \gamma_{ik}^{(t)} \propto \Pr[\bmx_i \cond z_i = k; \bTheta^{(t)}] \cdot \Pr[z_i = k \cond \bTheta^{(t)}] = \frac{\prod_{j=1}^n {(\alpha_{k}^{(t)})}^{\ind[x_{ij} = +1]} {(1-\alpha_{k}^{(t)})}^{\ind[x_{ij} = -1]} q_k^{(t)}}{\sum_{\ell=1}^K \prod_{j=1}^n {(\alpha_{\ell}^{(t)})}^{\ind[x_{ij} = +1]} {(1-\alpha_{\ell}^{(t)})}^{\ind[x_{ij} = -1]} q_{\ell}^{(t)}} \]
\item {\bf M-step}: Based on the current posterior estimates of the customer segment memberships $\gamma_{ik}^{(t)}$ and the observed data $\calD$, the model parameters are updated by maximizing $\mathbb{E} \bexpt{\log \Pr[\calD, \bm{z} \; | \; \bTheta^{(t)}]}$, which can be shown to be a lower bound on the data log-likelihood (eq \ref{eq:mle-em}). Equating the derivative of the expected conditional log-likelihood w.r.t $q_k$ to zero (with the additional constraint that $\sum_{\ell=1}^K q_{\ell} = 1$), we get the parameter estimate for the next iteration
\[ q_k^{(t+1)} = \frac{\sum_{i=1}^m \gamma_{ik}^{(t)}}{m} \; \; \; \; \; \text{for } k \in [K] \]
Similarly, for the parameters $\alpha_{k}$ we get the following expression
\[ \alpha_{k}^{(t+1)} = \frac{\sum_{i=1}^m \sum_{j=1}^n \gamma_{ik}^{(t)}\ind[x_{ij} = +1]}{\sum_{i=1}^m \sum_{j=1}^n \gamma_{ik}^{(t)}} \; \;\;\; \text{for }  k \in [K] \]
\end{itemize}
We repeat these two steps until convergence of the log-likelihood $\log \Pr[\calD \; | \; \bTheta]$.

%

\section{Simulation Details}
\label{app:impl}
We imposed $Beta(2,2)$ prior on the parameters $\alpha_k$ and $Dir(1.5,1.5,\ldots,1.5)$ prior on the parameters $q_k$ in the LC method, to avoid numerical issues for sparse graphs. Since the LC method is sensitive to the starting configuration, we ran it $10$ times with different random initializations and report the best outcome.

\end{APPENDICES}
\end{document}